\documentclass{aa} 

\usepackage{graphicx}
\usepackage{txfonts}

\usepackage{natbib}
\bibpunct{(}{)}{;}{a}{}{,} 

\begin{document}

\title{Multi-wavelength observations of Galactic hard X-ray sources
discovered by \textit{INTEGRAL}\thanks{Based on
observations carried out at the European Southern Observatory
under programmes ID 075.D-0773 and 077.D-0721.}}

\subtitle{II. The environment of the companion star}

\author{F. Rahoui
\inst{1,2}
\and
S. Chaty\inst{2}
\and
P-O. Lagage\inst{2}	
\and
E. Pantin\inst{2}
}

\offprints{F. Rahoui}

\institute{European Southern Observatory, Alonso de C\'ordova 3107,
Vitacura, Santiago de Chile\\
\email{frahoui@eso.org}
\and
Laboratoire AIM, CEA/DSM - CNRS - Universit\'e Paris Diderot,
IRFU/Service d'Astrophysique, B\^at. 709, CEA-Saclay, F-91191
Gif-sur- Yvette C\'edex, France\\
}

\date{Received ; accepted} 

\abstract
{The \textit{INTEGRAL} mission has led to the discovery of a new type of
	supergiant X-ray binaries (SGXBs), whose physical properties
	differ from those of previously known SGXBs. 	
	Those sources are in the course of being unveiled by means of
	multi-wavelength X-rays, optical, near- and mid-infrared
	observations, and two classes are appearing. 
	The first class consists of obscured
	persistent SGXBs and the second is populated by the
	so-called supergiant fast X-ray transients (SFXTs).}
{We report here mid-infrared (MIR) observations of the
	companion stars of twelve SGXBs 
	from these two classes in order to assess the
	contribution of the star and the material enshrouding the
	system to the total emission.}
{We used data from observations we carried out at ESO/VLT with
	VISIR, as well as
	archival and published data, to perform
	broad-band spectral energy distributions of the companion stars
	and fitted them with a combination of two black bodies representing
	the star and a MIR excess due to the absorbing material
	enshrouding the star, if there was any.}
{We detect a MIR excess in the emission of IGR~J16318-4848,
	IGR~J16358-4726, and perhaps IGR~J16195-4945. The other sources
	do not exhibit any MIR excess even when the intrinsic absorption is
	very high. Indeed, the
	stellar winds of supergiant stars are not suitable for dust
	production, and we show that this behaviour is not
	changed by the presence of the compact object. 
	Concerning IGR~J16318-4848 and
	probably IGR~J16358-4726, the MIR excess can
	be explained by their sgB[e] nature and the
	presence of an equatorial disk around the supergiant
	companion in which dust can be produced. Moreover, our results
	suggest that some of the supergiant stars in those systems could
	exhibit an absorption excess compared to isolated supergiant stars,
	this excess being possibly partly due to the
	photoionisation of their stellar wind in the vicinity of their
	atmosphere. (abridged)
}
{}

\titlerunning{MIR observations revealing the obscured
\textit{INTEGRAL} binary systems}

\maketitle
%

\section{Introduction}

High-mass X-ray binaries (HMXBs) are X-ray sources for which 
high-energy emission stems from accretion onto a compact object
(black hole or neutron star) of material coming from a massive companion
star. Until recently, the huge majority of known HMXBs
were Be/X-ray binaries, i.e a neutron star accreting from a disc
around a Be star. Most of these sources are transient, even if a
few are persistent weak X-ray emitters
($\textrm{L}_\textrm{x}\,\sim\,10^{34}$ erg~s$^{-{1}}$). The other
known HMXBs were supergiant X-ray binaries (SGXBs), composed of
a compact object orbiting around an early-type supergiant and fed by
accretion from the strong radiative wind of the companion. These
objects are persistent sources
($\textrm{L}_\textrm{x}\,\sim\,10^{36}$ erg~s$^{-{1}}$), and their
relative low number compared to the population of Be/X-ray
binaries was explained as the consequence of the short lifetime of
supergiant stars.

The launch of the \textit{INTErnational Gammay-Ray Astrophysics
Laboratory}
\citep[\textit{INTEGRAL}, ][]{2003Winkler} in October 2002 completely
changed the situation, as many more HMXBs whose companion stars
are supergiants were discovered during the monitoring
of the Galactic centre and the Galactic plane using the onboard
IBIS/ISGRI instruments \citep{2003Ubertini, 2003Lebrun}. Most of
these sources are reported in
\citet{2007Bird} and \citet{2007Bodaghee}, and their studies
have revealed two main features that were not present on previously
known SGXBs:

\begin{itemize} 
\item first, many of them exhibit a
	considerable intrinsic absorption, with a column density up to
	$N_{\rm H}\,\sim\,2\times10^{24}\,\textrm{cm}^{-2}$
in the case of IGR~J16318-4848
	\citep{2003Matt}, which explains why previous
	high-energy missions had not detected them.

\item second, some of these new sources reveal a transitory
nature. They are
	undetectable most of the time and occasionally present a fast
	X-Ray transient activity lasting a few hours. Moreover, they
	exhibit a quiescent luminosity of
	$\textrm{L}_\textrm{x}\,\sim\,10^{33}$ erg~s$^{-{1}}$, well below
	the persistent state of other SGXBs.
	\end{itemize}
It then appears that the supergiant HMXBs discovered by
\textit{INTEGRAL} can be classified in two classes: one
class of considerably obscured persistent sources that we will simply
call obscured SGXBs in this paper and another of
supergiant fast X-ray transients \citep[SFXTs,][]{2006Negueruelaa}.

High-energy observations can give some information about the
compact object or about the processes that lead to the emission
but do not allow study of the companion star. It is therefore
very important to perform multi-wavelength observations of these
sources - from optical-to-MIR wavelength - as this
represents the only way to characterise the
companion or to detect dust around these highly
obscured systems. However, positions
given by \textit{INTEGRAL} are
not accurate enough ($\sim$ 2\arcmin) to identify their optical
counterparts, because of the large number of objects in the
error circle. Observations with X-ray telescopes like
\textit{XMM-Newton} or \textit{Chandra} are therefore crucial
because they allow
a localisation with a position accuracy of 4$\arcsec$ 
or better, which lowers the number of possible optical
counterparts.

We performed optical-to-MIR wavelength observations of
several candidate SGXBs
recently discovered with \textit{INTEGRAL}. Optical and NIR
observations were carried out at ESO/NTT using EMMI
and SofI instruments and aimed at constraining 
the spectral type of the companions through accurate astrometry,
as well as the spectroscopy and photometry of the candidate
counterparts. They are reported in the companion
paper \citep[][ CHA08 hereafter]{2008Chaty}, and it is shown
that most of these sources are actually supergiant stars.

In this paper, we report MIR photometric observations of the companions
of twelve \textit{INTEGRAL} candidate SFXTs and obscured SGXBs
that aimed at studying the circumstellar environment of these highly
absorbed sources and, more particularly, at detecting any MIR excess in
their emission that could be due to the absorbing material. These
sources were chosen because they had very accurate positions
and confirmed 2MASS counterparts.

All the sources in our sample are listed in Table 1. The total
galactic column density of neutral hydrogen
$N_{\rm H}(\ion{H}{i})$ is computed using the web version
of the $N_{\rm H}$ FTOOL from HEASARC. This tool
uses the data from
\citet*{1990Dickey}, who performed \ion{H}{i} observations from
the Lyman-$\alpha$ and 21~cm lines. Moreover,
$N_{\rm H}(\ion{H}{i})$ is
the total galactic column density, which means it is
integrated along the line of sight over the whole
Galaxy. Therefore, it is likely to be overestimated compared to the
real value at the distance of the sources.

The total galactic column density of molecular hydrogen
$N_{\rm H}(\textrm{H}_{\rm 2})$ is computed using the
velocity-integrated map (W$_{\rm CO}$) and the X-ratio given in
\citet*{2001Dame}. It is also likely to be overestimated compared to the
real value at the distance of the sources, because it is
integrated along the line of sight over the whole Galaxy. In
contrast, $N_{\rm Hx}$, the intrinsic X-ray column density of the source, is computed
from the fitting of the high-energy spectral energy distribution
(SED), so it takes all the 
absorption into account at the right distance of the source.
\newline

Using these observations, the results reported
in the companion paper (CHA08), as well as archival
photometric data from the
USNO, 2MASS, and GLIMPSE
catalogues when needed, we performed the broad-band
SEDs of these sources and fitted them with a two-component
black body model to assess the 
contribution of the star and the enshrouding material in the
emission. The ESO observations, as well as our model, are described in
Section 2. In Section 3, results of the fits for each source
are given and these results are discussed in Section 4. We
conclude in Section 5. 


\section{Observations}

\subsection{MIR observations and data reduction}

The MIR observations were carried out on 2005 June 20-22 and 2006 June
29-30 using VISIR \citep{2004Lagage}, 
the ESO/VLT mid-infrared imager and spectrograph, composed of an
imager and a long-slit spectrometer covering several filters in
N and Q bands and
mounted on Unit 3 of the VLT (Melipal). The standard ''chopping
and nodding" MIR observational technique was used to suppress the
background dominating at these wavelengths. Secondary mirror-chopping was
performed in the north-south direction with an amplitude of
16$\arcsec$ at a frequency of 0.25 Hz. Nodding technique,
needed to compensate for chopping residuals, was chosen as
parallel to the chopping and applied
using telescope offsets of 16$\arcsec$. Because of the high
thermal MIR background for ground-based
observations, the detector integration
time was set to 16~ms.

We performed broad-band photometry in 3 filters, PAH1
($\lambda$=8.59$\pm$0.42 $\mu$m), PAH2
($\lambda$=11.25$\pm$0.59 $\mu$m), and Q2
($\lambda$=18.72$\pm$0.88 $\mu$m) using the small field in all bands 
(19\farcs2x19\farcs2 and 0\farcs075 plate scale). All
the observations were bracketed with standard star
observations for flux calibration and
PSF determination. The weather conditions
were good and stable during the observations.

Raw data were reduced using the IDL reduction package written by Eric
Pantin. The elementary images were co-added in real-time to obtain
chopping-corrected data, then the different nodding positions were
combined to form the final image. The VISIR detector is affected by
stripes randomly triggered by some abnormal high-gain pixels.
A dedicated destriping method was developed (Pantin 2008,
in prep.) to suppress them. The MIR fluxes of all observed
sources including the 1$\sigma$ errors are listed in Table 2.

\subsection{Archival data}

When we did not have optical-to-MIR data for our sources, we
searched for the counterparts in 3 catalogues: 
\begin{itemize}
\item in the United States Naval Observatory (USNO)
	catalogues in \textit{B}, \textit{R}, and \textit{I} for USNO-B1.0, \textit{B} and \textit{R}
	for USNO-A.2. Positions and fluxes
	accuracies are 0\farcs25 and 0.3 magnitudes in the case of
	USNO-B.1, 0\farcs2 and 0.5 magnitudes in the case of USNO-A.2.
\item in the 2 Micron All Sky Survey (2MASS), in \textit{J}
	(1.25$\pm$0.16 $\mu$m), \textit{H} (1.65$\pm$0.25 $\mu$m) and \textit{Ks}
	(2.17$\pm$0.26 $\mu$m) bands. Position accuracy is about 0\farcs2.
\item in the \textit{Spitzer}'s Galactic Legacy Infrared Mid-Plane Survey
	Extraordinaire \citep[GLIMPSE,][]{2003Benjamin}, survey of the
Galactic plane
	($|b|\,\leq\,1^\circ$ and between \textit{l}=10$^\circ$ and
\textit{l}=65$^\circ$ on both sides of the Galactic centre)
	performed with the \textit{Spitzer Space Telescope}, using the IRAC
	camera in four bands, 3.6$\pm$0.745 $\mu$m, 4.5$\pm$1.023 $\mu$m,
	5.8$\pm$1.450 $\mu$m, and 8$\pm$2.857 $\mu$m.
\end{itemize}
All sources had a confirmed 2MASS counterpart and three of them (IGR
J16195-4945, IGR~J16207-5129, and IGR~J16318-4848) had a GLIMPSE
counterpart given in the literature. We found all the other
GLIMPSE counterparts using the 2MASS positions and they are listed in
Table 3. We used all the
fluxes given in the GLIMPSE catalogue
except in the case of IGR~J17252-3616, IGR~J17391-3021 and
IGR~J17544-2619 because
their fluxes were not present in the catalogue tables. Nevertheless,
we measured their
fluxes on the archival images directly with aperture
photometry. Uncertainties on the
measurements were computed in the same way on the error
maps given with the data.

\subsection{Absorption}

Absorption at wavelength $\lambda$, $\textrm{A}_\lambda$, is a
crucial parameter to fit the SEDs, especially in the
MIR. Indeed, inappropriate values can lead to a false assessment
of the MIR excess. Visible
absorption A$_\textrm{v}$ was a free parameter of the fits. An
accurate interstellar absorption law - i.e. the wavelength
dependence of
the $\frac{\textrm{A}_\lambda}{\textrm{A}_\textrm{v}}$ ratio in
the line of sight - was then needed to properly fit the SEDs.

In the optical bands, we built the function with the analytical
expression given in
\citet{1989Cardelli} who derived the average
interstellar extinction law in the
direction of the Galactic centre. From 1.25 $\mu$m to 8 $\mu$m,
we used the analytical expression given in \citet[]{2005Indebetouw}. 
They derived it from the measurements of the mean
values of the colour excess ratios
$\frac{(\textrm{A}_\lambda-\textrm{A}_{\rm K})}
{(\textrm{A}_{\rm J} -\textrm{A}_{\rm K})}$
from the colour distributions of observed stars in the direction of the
Galactic centre. They used archival data from 2MASS
and GLIMPSE catalogues, which is relevant in our case as we use
GLIMPSE fluxes. 
\newline

Above 8 $\mu$m, where absorption is dominated by the silicate
features at 9.7 $\mu$m and 18 $\mu$m, we found several
extinction laws in the
literature \citep{1989Rieke, 1996Lutz, 2001Moneti}, which exhibit some
differences. Considering the high importance of a good assessment
to correctly fit the MIR excess, we decided to assess the ratio
$\frac{\textrm{A}_\lambda}{\textrm{A}_\textrm{v}}$ in 2
VISIR bands - PAH1 and PAH2 - from our data in order to
build the relevant law for our observations. 

\citet*{1985Rieke} gave the interstellar extinction law up to 13
$\mu{\textrm{m}}$, and from their results, we derived 
$0.043\leq\frac{\textrm{A}_{\rm PAH1}}{\textrm{A}_\textrm{v}}\leq0.074$
and
$0.047\leq\frac{\textrm{A}_{\rm PAH2}}{\textrm{A}_\textrm{v}}\,\leq0.06$.
To get the best values corresponding to our data in PAH1 and
PAH2, we proceeded
in 3 steps. 
\begin{itemize}
	\item We first selected the sources for which we had VISIR
fluxes in PAH1 and/or
	PAH2 and fitted their SEDs with extinction laws given in
	\citet{1989Cardelli} and \citet{2005Indebetouw} from 0.36 to 8 $\mu{\textrm{m}}$ and
	half-interval values taken in PAH1 and PAH2.
	\item Then, when we did not need any MIR excess to fit the
	IRAC fluxes, we adjusted the 
	$\frac{\textrm{A}_{\rm PAH1}}{\textrm{A}_\textrm{v}}$ and
	$\frac{\textrm{A}_{\rm PAH2}}{\textrm{A}_\textrm{v}}$ ratios to
	improve the $\chi^2$ of our fits.
	\item We finally averaged all the extinction values obtained for all
	sources to get what we consider as the right ratios in PAH1 and PAH2
	in the direction of the Galactic plane.
\end{itemize}
The resulting values are in good agreement with those given by
the extinction law from \citet{1996Lutz}, so we chose
their extinction law to fit our SEDs above 8
$\mu{\textrm{m}}$, the Q2 filter
included. The $\frac{\textrm{A}_\lambda}{\textrm{A}_\textrm{v}}$ values we
used in each band are listed in Table 4, and the overall
extinction law is displayed in Fig 1.

\subsection{SEDs}

With all the archival and observational data from optical-to-MIR 
wavelength, we built the SEDs for these sources. We fitted them
(using a $\chi^2$ minimisation) with a
model combining two absorbed black bodies, one representing the companion
star emission and a spherical one representing a possible MIR
excess due to the absorbing material enshrouding the companion star:

\begin{displaymath}
	\lambda{F(\lambda)}\,=\,\frac{2\pi{h}{c}^2}{{D_{\ast}}^2{\lambda}^4}
\,10^{\textrm{\normalsize
	$-{0.4A_\lambda}$}}\left[\frac{{R_\ast}^2}{
	e^{\textrm{\large
	$\frac{hc}{{\lambda}k{T}_\ast}$}}-1}+\frac{{R_{\rm D}}^2}{e^{\textrm{\large
	$\frac{hc}{{\lambda}k{T}_{\rm D}}$}}-1}\right]\,\,\,\,\,\,\,\textrm{in
	W m}^{-2}
\end{displaymath}

We added to the flux uncertainties systematic errors as follows:

\begin{itemize}
	\item a 2$\%$ systematic error in each IRAC band as given in
the IRAC
manual\footnote{http://ssc.spitzer.caltech.edu/documents/som/
som8.0.irac.pdf}
	\item comparing the variations of the flux calibration
values obtained from standards with VISIR during our
observation nights, we figured out that systematic errors
with VISIR were about 5$\%$ at 10 $\mu{\textrm{m}}$ and
10$\%$ at 20 $\mu{\textrm{m}}$.
\end{itemize}
The free parameters of the fits were the absorption in the
V-band A$_\textrm{v}$,
the companion star black body temperature T$_\ast$ and radius
to distance ratio $\frac{\textrm{R}_\ast}{\textrm{D}_{\ast}}$, as
well as the additional spherical component black body temperature
and radius T$_{\rm D}$ and $\textrm{R}_{\rm D}$.\\	
\newline
The best-fitting
parameters for individual sources, as well as corresponding
$\chi^2$ are
listed in Table 5 and the fitted SEDs are displayed in Fig 3.
Moreover, 90\%-confidence ranges of parameters are listed in
Table 6. 

In Table 5, along with the best-fitting parameters, we also give
the total galactic 
extinctions in magnitudes $\textrm{A}_{\ion{H}{i}}$ and
$\textrm{A}_{\rm H_2}$ in the line of sight, as well as the
X-ray extinction of the source
in magnitudes $\textrm{A}_\textrm{x}$. The values of
$\textrm{A}_{\ion{H}{i}}$, $\textrm{A}_{\rm H_2}$, and
$\textrm{A}_\textrm{x}$ are computed from
$N_{\rm H}(\ion{H}{i})$, 
$N_{{\rm H}}(\textrm{H}_{\rm 2})$, and
$N_{\rm Hx}$ given in Table 1 using the relation 
$\textrm{A}_{\rm H}\,=\,\frac{3.1}{5.8\times10^{21}\,\textrm{cm}^{-{2
}}}N_{\rm H}$
\citep*{1978Bohlin,1985Rieke}.

\begin{table*}
	\caption{Sample of sources studied in this paper. 
	We give their name, their
	coordinates (J2000 and galactic), the total galactic
	column density of neutral hydrogen
	($N_{\rm H}(\ion{H}{i})$) and the total galactic
	column density of molecular hydrogen
($N_{\rm H}(\textrm{H}_{\rm 2})$) in the line of
	sight, the X-ray column
	density of the source ($N_{\rm Hx}$), their type
(SFXT or OBS - obscured sources) and their
	spectral type (SpT). Their spectral classifications come from optical/NIR
	spectroscopy, reported in the following references (Ref): c:
\citet{2008Chaty}, f: \citet*{2004Filliatre}, i: \citet{2006Zand},
	n1: \citet{2005Negueruela}, n2: \citet{2006Negueruelab}, n3:
\citet{2007Nespoli}, p:
	\citet{2006Pellizza}, t: \citet{2006Tomsick}.}
	$$
	\begin{array}{c c c c c c c c c c c}
	\hline
	\hline
	\textrm{Sources}&\alpha (\textrm{J2000})&\delta
(\textrm{J2000})&l&b&N_{\rm H}(\ion{H}{i})(10^{22})&
N_{\rm H}(\textrm{H}_{\rm 2})(10^{22})&N_{\rm Hx}
(10^{22})&\textrm{Type}&\textrm{SpT}&\textrm{Ref}\\
	\hline
	\textrm{IGR~J16195-4945}&16\,\,19\,\,32.20&-49\,\,44\,\,30.7&3
33.56&0.339&2.2&5.4&7.0&\textrm{SFXT ?}&\textrm{O/B}&\textrm{t}\\
	\hline
	\textrm{IGR~J16207-5129}&16\,\,20\,\,46.26&-51\,\,30\,\,06.0&
332.46&-1.050&1.7&2.2&3.7&\textrm{OBS}&\textrm{O/B}&\textrm{t}\\
	\hline
	\textrm{IGR~J16318-4848}&16\,\,31\,\,48.60&-48\,\,49\,\,00.0&
335.62&-0.448&2.1&3.6&200.0&\textrm{OBS}&\textrm{sgB[e]}&\textrm{f}\\
	\hline
	\textrm{IGR~J16320-4751}&16\,\,32\,\,01.90&-47\,\,52\,\,27.0&
336.30&0.169&2.1&4.4&21.0&\textrm{OBS}&\textrm{O/BI}&\textrm{c}\\
	\hline
	\textrm{IGR~J16358-4726}&16\,\,35\,\,53.80&-47\,\,25\,\,41.1&
337.01&-0.007&2.2&7.3&33.0&\textrm{OBS}&\textrm{sgB[e]}&\textrm{c}\\
	\hline
	\textrm{IGR~J16418-4532}&16\,\,41\,\,51.00&-45\,\,32\,\,25.0&
339.19&0.489&1.9&3.6&10.0&\textrm{SFXT ?}&\textrm{O/B}&\textrm{c}\\
	\hline
	\textrm{IGR~J16465-4507}&16\,\,46\,\,35.50&-45\,\,07\,\,04.0
&340.05&0.135&2.1&5.9&60.0&\textrm{SFXT}&\textrm{B0.5I}&\textrm{n1}\\
	\hline
	\textrm{IGR~J16479-4514}&16\,\,48\,\,06.60&-45\,\,12\,\,08.0&
340.16&-0.124&2.1&8.2&7.7&\textrm{SFXT ?}&\textrm{O/BI}&\textrm{c}\\
	\hline
	\textrm{IGR~J17252-3616}&17\,\,25\,\,11.40&-36\,\,16\,\,58.6&
351.50&-0.354&1.6&3.9&15.0&\textrm{OBS}&\textrm{O/BI}&\textrm{c}\\
	\hline
	\textrm{IGR~J17391-3021}&17\,\,39\,\,11.58&-30\,\,20\,\,37.6&
358.07&0.445&1.4&4.5&30.0&\textrm{SFXT}&\textrm{08Iab(f)}&\textrm{n2}\\
	\hline
	\textrm{IGR~J17544-2619}&17\,\,54\,\,25.28&-26\,\,19\,\,52.6
&3.26&-0.336&1.4&7.8&1.4&\textrm{SFXT}&\textrm{O9Ib}&\textrm{p}\\
	\hline
	\textrm{IGR~J19140+0951}&19\,\,14\,\,04.23&+09\,\,52\,\,58.3&
44.30&-0.469&1.7&3.7&6.0&\textrm{OBS}&\textrm{B1I}&\textrm{n3}\\
	\hline
	\end{array}
	$$
\end{table*}

\begin{table*}
	\caption{Summary of VISIR observations of newly discovered
\textit{INTEGRAL}
	sources. We give their MIR fluxes (mJy) in the PAH1 (8.59 $\mu$m), PAH2
(11.25 $\mu$m) and
	Q2 (18.72 $\mu$m) filters. When we did not detect a source,
	we give the upper limit. When no flux nor upper limit is
	given, the source was not observed in the considered filter.}
	$$
	\begin{array}{c c c c}
	\hline
	\hline
	\textrm{Sources}&\textrm{PAH1}&\textrm{PAH2}&\textrm{Q2}\\
	\hline
	\textrm{IGR~J16195-4945}&<6.1&<7.8&<50.3\\
	\hline
	\textrm{IGR~J16207-5129}&21.7\pm1.4&9.4\pm1.3&<53.4\\
	\hline
	\textrm{IGR~J16318-4848}&426.2\pm3.0&317.4\pm3.4&180.7\pm15.3\\
	\hline
	\textrm{IGR~J16320-4751}&12.1\pm1.7&6.3\pm1.8&\\
	\hline
	\textrm{IGR~J16358-4726}&<6.9&&\\
	\hline
	\textrm{IGR~J16418-4532}&<5.8&&\\
	\hline
	\textrm{IGR~J16465-4507}&6.9\pm1.1&<5.0&\\
	\hline
	\textrm{IGR~J16479-4514}&10.9\pm1.2&7.0\pm1.6&\\
	\hline
	\textrm{IGR~J17252-3616}&6.1\pm0.6&<5.0&\\
	\hline
	\textrm{IGR~J17391-3021}&70.2\pm1.6&46.5\pm2.6&\\
	\hline
	\textrm{IGR~J17544-2619}&46.1\pm2.8&20.2\pm2.1&\\
	\hline
	\textrm{IGR~J19140+0951}&35.2\pm1.4&19.1\pm1.4&\\
	\hline
	\end{array}
	$$
\end{table*}

\begin{table*}
	\caption{List of GLIMPSE counterparts we found for 9
sources. We give their name, their separation from the 2MASS counterparts
	and their fluxes in mJy.}
	$$
	\begin{array}{c c c c c c c}
	\hline
	\hline
	\textrm{Sources}&\textrm{GLIMPSE
counterpart}&\textrm{Separation}&3.6\,\mu\textrm{m}&4.5\,\mu
\textrm{m}&5.8\,\mu\textrm{m}&8\,\mu\textrm{m}\\
	\hline
	\textrm{IGR~J16320-4751}&\textrm{G336.3293+00.1689}&0\farcs17&
48.2\pm1.9&44.3\pm2.1&36.0\pm2.0&17.3\pm1.0\\
	\hline
	\textrm{IGR~J16358-4726}&\textrm{G337.0994-00.0062}&0\farcs46&
5.9\pm0.5&5.6\pm0.6&5.3\pm1.7&\\
	\hline
	\textrm{IGR~J16418-4532}&\textrm{G339.1889+00.4889}&0\farcs28&
12.5\pm0.9&9.5\pm0.6&5.6\pm0.6&3.6\pm0.4\\
	\hline
	\textrm{IGR~J16465-4507}&\textrm{G340.0536+00.1350}&0\farcs16&
45.0\pm2.0&32.6\pm1.5&22.0\pm0.9&13.7\pm0.6\\
	\hline
	\textrm{IGR~J16479-4514}&\textrm{G340.1630-00.1239}&0\farcs13&
68.6\pm3.1&49.6\pm2.0&41.2\pm2.3&19.4\pm0.9\\
	\hline
	\textrm{IGR~J17252-3616}&&&32.6\pm3.7&24.7\pm4.7&21.8\pm5.0&
9.6\pm6.5\\
	\hline
	\textrm{IGR~J17391-3021}&&&375.9\pm44.0&297.0\pm31.0&205.0\pm33.0&
111.0\pm28.0\\
	\hline
	\textrm{IGR~J17544-2619}&&&213.9\pm25.4&137.0\pm18.9&99.6\pm7.6&
66.5\pm12.1\\
	\hline
	\textrm{IGR~J19140+0951}&\textrm{G044.2963-00.4688}&0\farcs48&
185.0\pm9.3&152.0\pm11.1&103.9\pm5.4&62.0\pm2.1\\
	\hline
	\end{array} 
	$$
\end{table*} 

\begin{table*}
	\caption{Adopted $\frac{\textrm{A}_{\lambda}}{\textrm{A}_{v}}$ values.}
	$$
	\begin{array}{c c c c c c c c c c c c c c c c}
	\hline
	\hline
	\textrm{Filters}&\textrm{\textit{U}}&\textrm{\textit{B}}&\textrm{\textit{V}}&\textrm{\textit{R}}&
\textrm{\textit{I}}&\textrm{\textit{J}}&\textrm{\textit{H}}&\textrm{\textit{Ks}}&3.6\,\mu\textrm{m}&
4.5\,\mu\textrm{m}&5.8\,\mu\textrm{m}&8\,\mu\textrm{m}&8.59\,\mu
\textrm{m}&11.25\,\mu\textrm{m}&18.72\,\mu\textrm{m}\\
	\hline
	\frac{\textrm{A}_{\lambda}}{\textrm{A}_{v}}&1.575&1.332&1&0.757&
0.486&0.289&0.174&0.115&0.0638&0.0539&0.0474&0.0444&0.0595&0.0605&
0.040\\
	\hline
	\end{array} 
	$$
\end{table*}

\begin{figure*}
	\centering
	\includegraphics[angle=270,width=10cm]{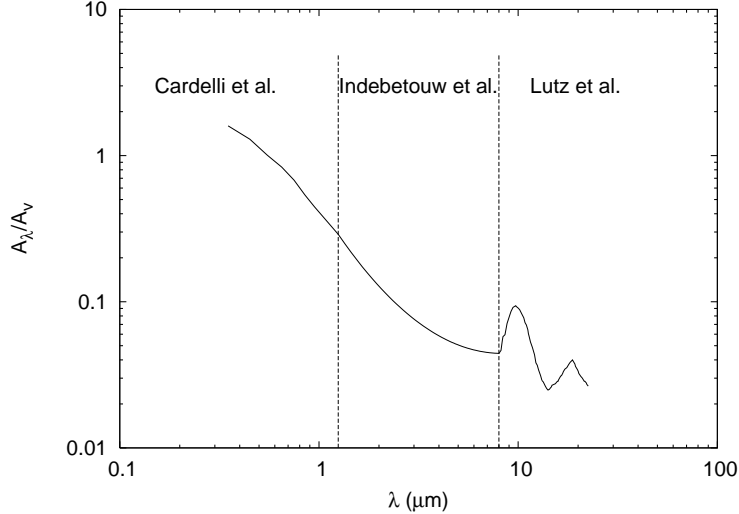}
	\caption{Adopted extinction law. We used the law given in
	\citet{1989Cardelli} in the optical, the one given in
	\citet{2005Indebetouw} from 1.25 $\mu{\textrm{m}}$ to 8
$\mu{\textrm{m}}$, and the law
	from \citet{1996Lutz} above 8 $\mu{\textrm{m}}$.}
\end{figure*}

\begin{table*}
	\caption{Summary of best-fitting parameters of the SEDs of the
	sources. We give the total galactic 
	extinctions in magnitudes $\textrm{A}_{\ion{H}{i}}$ and
	$\textrm{A}_{\rm H_2}$, the X-ray extinction of the source
	in magnitudes $\textrm{A}_\textrm{x}$ and then the
	parameters themselves: the extinction in the optical
	$\textrm{A}_\textrm{v}$, the temperature
	$\textrm{T}_\ast$ and the
	$\frac{\textrm{R}_{\ast}}{\textrm{D}_{\ast}}$ ratio of the
	companion and the
	temperature and radius $\textrm{T}_{\rm D}$ and
$\textrm{R}_{{\rm D}}$ (in
	$\textrm{R}_{\ast}$ unit) of the dust component
	when needed. We also add the reduced $\chi^2$ we reach for
	each fit.} 
	$$
	\begin{array}{c c c c c c c c c c}
	\hline
	\hline
	\textrm{Sources}&\textrm{A}_{\ion{H}{i}}&\textrm{A}_{\rm H_2}&
\textrm{A}_\textrm{x}&\textrm{A}_\textrm{v}&\textrm{T}_{\ast}
(\textrm{K})&\frac{\textrm{R}_{\ast}}{\textrm{D}_{\ast}}&\textrm{T}_
{\rm D}
(\textrm{K})&\textrm{R}_{\rm D}
\left(\textrm{R}_{\ast}\right)&\chi^2/\textrm
{dof}\\
	\hline
	\textrm{IGR~J16195-4945}&11.7&28.6&37.4&15.5&23800&5.96\times
10^{-{11}}&1160&5.1&3.9/2\\
	\hline
	\textrm{IGR~J16207-5129}&9.3&11.8&19.8&10.5&33800&9.42\times
10^{-{11}}&&&28.5/9\\
	\hline
	\textrm{IGR~J16318-4848}&11.0&19.2&1069.5&17.0&22200&3.74\times
10^{-{10}}&1100&10.0&6.6/6\\
	\hline
	\textrm{IGR~J16320-4751}&11.4&23.6&112.2&35.4&33000&1.38\times
10^{-{10}}&&&7.7/6\\
	\hline
	\textrm{IGR~J16358-4726}&11.8&39.4&176.4&17.6&24500&3.16\times
10^{-{11}}&810&10.1&3.6/2\\
	\hline
	\textrm{IGR~J16418-4532}&10.1&19.3&53.5&14.5&32800&3.77\times
10^{-{11}}&&&1.4/4\\
	\hline
	\textrm{IGR~J16465-4507}&11.3&31.4&320.7&5.9&25000&6.40\times
10^{-{11}}&&&13.9/7\\
	\hline
	\textrm{IGR~J16479-4514}&11.4&43.8&41.2&18.5&32800&1.00\times
10^{-{10}}&&&7.4/6\\
	\hline
	\textrm{IGR~J17252-3616}&8.3&20.1&80.2&20.8&32600&7.57\times
10^{-{11}}&&&3.8/5\\
	\hline
	\textrm{IGR~J17391-3021}&7.3&23.9&160.4&9.2&31400&1.80\times
10^{-{10}}&&&11.7/10\\
	\hline
	\textrm{IGR~J17544-2619}&7.7&41.5&7.70&6.1&31000&1.27\times10^{-{10}}
&&&6.1/8\\
	\hline
	\textrm{IGR~J19140+0951}&9.0&14.0&32.1&16.5&22500&1.92\times10^
{-{10}}&&&14.4/6\\
	\hline
	\end{array}
	$$
\end{table*} 

\begin{table*}
	\caption{Ranges of parameters that give acceptable
	fits (90\%-confidence) for each source.
	}
	$$
	\begin{array}{c c c c c c}
	\hline
	\hline
	\textrm{Sources}&\Delta
	\textrm{A}_\textrm{v}&\Delta\textrm{T}_{\ast}&\Delta\frac{\textrm{R
}_{\ast}}{\textrm{D}_{\ast}}&\Delta\textrm{T}_{\rm D}&\Delta\textrm
{R}_{{\rm D}}\\
	\hline
	\textrm{IGR~J16195-4945}&14.8-15.8&13100-25900&5.68\times10^{-{11}}
-6.68\times10^{-{11}}&950-1460&3.9-6.6\\
	\hline
	\textrm{IGR~J16207-5129}&10.4-10.6&25200-36000&9.36\times10^{-{11}}
-1.05\times10^{-{10}}&&\\
	\hline
	\textrm{IGR~J16318-4848}&16.6-17.4&19300-24500&3.65\times10^{-{10}}
-3.84\times10^{-{10}}&960-1260&8.8-11.8\\
	\hline
	\textrm{IGR~J16320-4751}&34.8-35.5&22000-35600&1.33\times10^{-{10}}
-1.69\times10^{-{10}}&&\\
	\hline
	\textrm{IGR~J16358-4726}&17.1-18.1&19500-36000&2.64\times10^{-{11}}
-3.52\times10^{-{11}}&630-1020&8.0-13.8\\
	\hline
	\textrm{IGR~J16418-4532}&13.6-14.7&10600-36000&3.58\times10^{-{11}}
-5.44\times10^{-{11}}&&\\
	\hline
	\textrm{IGR~J16465-4507}&5.0-6.1&15400-33600&5.50\times10^{-{11}}
-7.95\times10^{-{11}}&&\\
	\hline
	\textrm{IGR~J16479-4514}&18.4-18.8&26200-36000&9.48\times10^{-{11}}
-1.13\times10^{-{10}}&&\\
	\hline
	\textrm{IGR~J17252-3616}&20.3-21.0&20500-36000&7.17\times10^{-{11}}
-9.48\times10^{-{11}}&&\\
	\hline
	\textrm{IGR~J17391-3021}&8.8-9.4&16100-32200&1.78\times10^{-{10}}
-2.63\times10^{-{10}}&&\\
	\hline
	\textrm{IGR~J17544-2619}&6.0-6.2&26700-35500&1.18\times10^{-{10}}
-1.38\times10^{-{10}}&&\\
	\hline
	\textrm{IGR~J19140+0951}&15.7-16.7&13200-28100&1.71\times10^{-{10}}
-2.53\times10^{-{10}}&&\\
	\hline
	\end{array}
	$$
\end{table*}

\section{Results}

\subsection{IGR~J16195-4945}

IGR~J16195-4945 was detected by \textit{INTEGRAL} during
observations carried
out between 2003 February 27 and October 19 \citep{2004Walter} and
corresponds to the ASCA source AX J161929-4945
\citep{2001Sugizaki,2005Sidoli}. 
Analysing \textit{INTEGRAL} public data,
\citet{2005Sidoli} derive an average flux level of
$\sim$ 17~mCrab (20-40~keV). Performing a follow-up with
\textit{INTEGRAL}, \citet{2006Sguera} show it behaves like
an SFXT and report a peak-flux of $\sim$ 35~mCrab (20-40~keV).

\citet{2006Tomsick} observed the source with \textit{Chandra}
between 2005 April and
July and give its position with 0\farcs6 accuracy. They fitted its
high-energy emission with an absorbed
power law and derive $\Gamma\sim0.5$ and
$N_{\rm H}\sim7\times10^{22}$~cm$^{-2}$. Moreover,
using their accurate localisation, they found its NIR and MIR
counterparts in
the 2MASS (2MASS~J16193220-4944305) and in the GLIMPSE
(G333.5571+00.3390) catalogues and performed its NIR photometry
using ESO/NTT observations. They show its spectral type is
compatible with an O, B, or A supergiant star. They also found possible
USNO-A.2 and USNO-B.1 counterparts. Nevertheless, as already
suggested in their paper, the USNO source is
a blended foreground object \citep{2006Tovmassian}. 
\newline

We observed IGR~J16195-4945 on 2006 June 30 in PAH1 during 1200~s, but
did not detect it. Typical seeing and airmass were
0\farcs88\ and 1.07. We nevertheless fitted its SED
using the NIR and the GLIMPSE flux values given in \citet{2006Tomsick}
and the best-fitting parameters are $\textrm{A}_\textrm{v}$=15.5,
$\textrm{T}_*$=23800~K,
$\frac{\textrm{R}_\ast}{\textrm{D}_{\ast}}$=$5.96\times10^{-{11}}$,
$\textrm{T}_{\rm D}$=1160~K,
$\textrm{R}_{\rm D}$=5.1$\textrm{R}_\ast$, and the reduced
$\chi^2$ is 3.9/2. 
\newline
The best-fitting parameters without the additional component are
$\textrm{A}_\textrm{v}$=16.1,
$\textrm{T}_*$=13800~K,
$\frac{\textrm{R}_\ast}{\textrm{D}_{\ast}}$=$8.94\times10^{-{11}}$,
and the corresponding reduced $\chi^2$ is 15.8/4. We then found
a MIR excess.

The additional component is then needed to correctly fit the SED, as
this source exhibits a MIR excess. Nevertheless, as shown in
Fig 4, this excess is small, and the lack of data above 8
$\mu$m does not allow to reach definitive conclusions. Moreover,
in both cases (with and without dust), the stellar
component is consistent with an O/B supergiant, as already
suggested in \citet{2006Tomsick}. 

\subsection{IGR~J16207-5129}

IGR~J16207-5129 is an obscured SGXB that was discovered by
\textit{INTEGRAL} during observations carried
out between 2003 February 27 and October 19 \citep{2004Walter}.

\citet{2006Tomsick} observed it with \textit{Chandra} during the
same run as
IGR~J16195-4945 and give its position with 0\farcs6
accuracy. 
They also fitted its high-energy emission with an absorbed
power law and derive $\Gamma\sim0.5$ and
$N_{\rm H}\sim3.7\times10^{22}$~cm$^{-2}$.
Thanks to their accurate localisation, they found its NIR and MIR
counterparts in
the 2MASS (2MASS~J16204627-5130060) and in the GLIMPSE
(G333.4590+01.0501) catalogues and performed its NIR photometry
using ESO/NTT observations. They show its temperature to
be $\geqslant$18000~K, which indicates the system is an
HMXB. They also found its USNO-B1.0
(USNO-B1.0~0384-0560875) counterpart.
\newline

We observed IGR~J16207-5129 on 2006 June 29 in PAH1 and PAH2 during
1200~s in each filter, and in Q2 during 2400~s. Typical seeing and
airmass were 0\farcs72 and 1.09. We did not detect it in Q2 but in PAH1
and PAH2. The fluxes we derived are 21.7$\pm$1.4~mJy and
9.4$\pm$1.3~mJy, respectively.

Using those values, as well as the fluxes from ESO/NTT observations and
the GLIMPSE archives found in \citet{2006Tomsick}, we fitted its SED,
and the best-fitting parameters are $\textrm{A}_\textrm{v}$=10.5,
$\textrm{T}_*$=33800~K,
$\frac{\textrm{R}_\ast}{\textrm{D}_{\ast}}$=$9.42\times10^{-{11}}$,
and the reduced $\chi^2$ is
28.5/9. \citet*{2007Negueruela} find the spectral type is earlier
than B1I; our parameters are therefore in good agreement with
their results.

The best fit with the additional component gives a larger
reduced $\chi^2$ of 30/7
and $\textrm{T}_{\rm D}\,<\,200$ K, which is not 
significant, as the presence of such cold material marginally enhances
the MIR flux. We therefore think IGR~J16207-5129 is an O/B
massive star whose enshrouding material marginally contributes
to its MIR emission.

\subsection{IGR~J16318-4848}

Main high-energy characteristics of this source can be found in
\citet*{2003Matt} and \citet{2003Walter}. IGR~J16318-4848 was
discovered by
\textit{INTEGRAL} on 2003 January 29 \citep{2003Courvoisier} and
was then
observed with \textit{XMM-Newton}, which allowed a 4$\arcsec$ localisation. Those
observations showed that the source was exhibiting a strong absorption
of $N_{\rm H}\sim2\times10^{24}$~cm$^{-2}$, a temperature kT = 9~keV, and a photon index $\sim\,2$. 

Using this accurate position, \citet*{2004Filliatre} discovered its
optical counterpart and confirmed the NIR counterpart proposed by
\citet{2003Walter} (2MASS~J16314831-4849005). They also performed
photometry and spectroscopy in optical and NIR on 2003 February 23-25
at ESO/NTT and show that the source presents a significant NIR excess
and that it is strongly absorbed
($\textrm{A}_\textrm{v}\sim17.4$). The spectroscopy revealed an
unusual spectrum with a continuum very rich in strong emission
lines, which, together with the presence of forbidden lines,
points towards an sgB[e] companion star \citep[see e.g. ][ for
definition and characteristics of these stars]{1998Lamers,1999Zickgraf}.

Using the 2MASS magnitudes, the GLIMPSE (G335.6260-00.4477), and
the MSX fluxes,
\citet{2006Kaplan} fitted its SED with a combination of a stellar
and a dust component black bodies, and shows that the presence
of warm dust around the system was necessary for explaining the
NIR and MIR excess. From their fit, they derive 
$\textrm{A}_\textrm{v}\sim18.5$, $\textrm{T}_{\rm D}$=1030~K,
and $\textrm{R}_{\rm D}$=10$\textrm{R}_{\ast}$.
\newline

We observed IGR~J16318-4848 with VISIR twice: 
\begin{itemize}
\item the first time on 2005
	June 21 during 300~s in PAH1 and PAH2, and 600~s in Q2. Typical seeing
	and airmass were 0\farcs81 and 1.14. We detected the source
	in all bands, and the derived fluxes are
	409.2$\pm$2.4~mJy, 322.4$\pm$3.3~mJy, and 172.1$\pm$14.9~mJy
in PAH1, PAH2, and Q2, respectively.
\item the second on 2006 June 30 during 600~s in all bands. Typical seeing
	and airmass were 0\farcs68 and 1.09. We 
	detected the source in all bands, and the derived fluxes are
	426.2$\pm$3.0~mJy, 317.4$\pm$3.4~mJy, and 180.7$\pm$15.3~mJy
in PAH1, PAH2, and Q2, respectively.
\end{itemize}
Those observations show that IGR~J16318-4848 is very bright in
the MIR (it is actually the brightest source in our sample) and
that its flux was constant within a year, considering VISIR
systematic errors.

Using data from our last run, as well as the magnitudes given in
\citet*{2004Filliatre} in the optical and the NIR, and the fluxes from
the GLIMPSE archives, we fitted its SED
and the best-fitting parameters are $\textrm{A}_\textrm{v}$=17,
$\textrm{T}_*$=22200~K,
$\frac{\textrm{R}_\ast}{\textrm{D}_{\ast}}$=$3.74\times10^{-{10}}$,
$\textrm{T}_{\rm D}$=1100~K,
$\textrm{R}_{\rm D}$=10$\textrm{R}_\ast$, and the reduced
$\chi^2$ is 6.6/6.
The best-fitting parameters without the additional component are
$\textrm{A}_\textrm{v}$=17.9,
$\textrm{T}_*$=18200~K,
$\frac{\textrm{R}_\ast}{\textrm{D}_{\ast}}$=$5.1\times10^{-{10}}$,
and the corresponding reduced $\chi^2$ is 425/8. We then confirm
that the MIR excess is likely due to the 
presence of warm dust around the system, as already suggested by
\citet*{2004Filliatre} and reported in \citet{2006Kaplan}.

\subsection{IGR~J16320-4751}

IGR~J16320-4751 was detected by \textit{INTEGRAL} on 2003
February \citep{2003Tomsick} 
and corresponds to the ASCA source AX
J1631.9-4752. \citet{2003Rodrigueza} report observations with
\textit{XMM-Newton}. They give an accurate localisation (3$\arcsec$) and
fitted its high-energy spectrum with an absorbed power law. They
derive $\Gamma\sim1.6$ and
$N_{\rm H}\sim2.1\times10^{23}$~cm$^{-2}$. 

\citet{2005Lutovinov} report the discovery of X-Ray pulsations
(P$\sim1309$~ s), which proves the compact object is a neutron
star. Moreover, \citet{2004Corbet} obtained the light curve of IGR
J16320-4751 between 2004 December 21 and 2005 September 17 with Swift
and report the discovery of a 8.96 days orbital period. IGR
J16320-4751 is then an X-ray binary whose compact object is a
neutron star.

\citet*{2007Negueruela} searched for the NIR counterpart of the
source 
in the 2MASS catalogue and found its position was consistent with
2MASS~J16320215-4752289. They also concluded that, if it was an O/B
supergiant, it had to be extremely absorbed. 

The optical and NIR photometry and spectroscopy of this source were
carried out at ESO/NTT, and results are reported in CHA08. It is
shown that its NIR spectrum is consistent with an O/B supergiant
and that its intrinsic absorption is very high, because it was not
detected in any of the visible bands. We searched for the MIR
counterpart of IGR~J16320-4751 in the GLIMPSE archives and found it
to be consistent with G336.3293+00.1689.
\newline

We observed IGR~J16320-4751 with VISIR on 2005 June 20 in PAH1 and
PAH2, and the respective exposure
times were 1800~s and 2400~s. Typical seeing and
airmass were 0\farcs63 and 1.13. We detected it in both filters, and
the respective fluxes are 12.1$\pm$1.7~mJy and 6.3$\pm$1.8~mJy.
Using the ESO/NTT NIR magnitudes given in CHA08, as well as the
GLIMPSE and the VISIR fluxes, we fitted its SED and the best-fitting
parameters are $\textrm{A}_\textrm{v}$=35.4,
$\textrm{T}_*$=33000~K,
$\frac{\textrm{R}_\ast}{\textrm{D}_{\ast}}$=$1.38\times10^{-{10}}$,
and the reduced $\chi^2$ is
7.7/6. This result is in good agreement with an extremely
absorbed O/B supergiant as reported in CHA08.

The best fit with the additional component gives a larger
reduced $\chi^2$ of 8/4
and $\textrm{T}_{\rm D}\,<\,200$ K. We therefore
think that IGR~J16320-4751 is an O/B
supergiant whose enshrouding material marginally contributes to
its MIR emission, even if its intrinsic absorption is extremely high.

\subsection{IGR~J16358-4726}

IGR~J16358-4726 was detected with \textit{INTEGRAL} on 2003
March 19 \citep{2003Revnivtsev} and first observed with
\textit{Chandra} on 2003 March 24 
\citep{2004Patel}. They give its position with 0\farcs6 accuracy
and fitted its high-energy spectrum with an absorbed power law. They
derive $\Gamma\sim0.5$ and
$N_{\rm H}\sim3.3\times10^{23}$~cm$^{-2}$. They also found a
5880$\pm$50~s modulation, which could be either a neutron star
pulsation or an orbital modulation. Nevertheless, \citet{2006Patel}
performed detailed spectral and timing analysis of this source using
multi-satellite archival observations and identified a 94~s spin up,
which points to a neutron star origin. Assuming that this spin up was due
to accretion, they estimate the source magnetic field is between $10^{13}$
and $10^{15}$ G, which could support a magnetar nature for
IGR~J16358-4726. 

\citet{2003Kouveliotou} propose 2MASS~J16355369-4725398 as the
possible NIR counterpart, and NIR spectroscopy and photometry of
this counterpart was performed 
at ESO/NTT and is reported in CHA08. They show that its
spectrum is consistent with a B supergiant belonging to the
same family as IGR~J16318-4848, the so-called B[e]
supergiants. We also found its MIR counterpart in the GLIMPSE
archives (G337.0994-00.0062).
\newline

We observed IGR~J16358-4726 with VISIR on 2006 June 29 but did
not detect it in any filter. Using the NIR magnitudes given in CHA08
and the GLIMPSE fluxes, we fitted its SED
and the best-fitting parameters are $\textrm{A}_\textrm{v}$=17.6,
$\textrm{T}_*$=24500~K,
$\frac{\textrm{R}_\ast}{\textrm{D}_{\ast}}$=$3.16\times10^{-{11}}$,
$\textrm{T}_{\rm D}$=810~K,
$\textrm{R}_{\rm D}$=10.1$\textrm{R}_\ast$, and the reduced
$\chi^2$ is 3.6/2. The best-fitting parameters
without the additional component are $\textrm{A}_\textrm{v}$=16.7,
$\textrm{T}_*$=9800~K,
$\frac{\textrm{R}_\ast}{\textrm{D}_{\ast}}$=$6.05\times10^{-{11}}$
and the corresponding reduced $\chi^2$ is 8.8/4. 

The additional component is then necessary to correctly fit the
SED, since
this source exhibits a MIR excess (see Fig 4). Even if we lack
MIR data above 5.8 $\mu$m, we think this excess is real and
stems from warm dust, as it
is consistent with the source being a sgB[e], as reported in CHA08. 

\subsection{IGR~J16418-4532}

IGR~J16418-4532 was discovered with \textit{INTEGRAL} on 2003
February 1-5
\citep{2004Tomsick}. Using \textit{INTEGRAL} observations,
\citet{2006Sguera} report an SFXT behaviour of this source and a
peak-flux of $\sim\,$80~mCrab (20-30~keV). Moreover, using
\textit{XMM-Newton} and
\textit{INTEGRAL} observations, \citet{2006Walter} report a
pulse period of 1246$\pm$100~s
and derive $N_{\rm H}\sim10^{23}$~cm$^{-2}$. They also
proposed 2MASS~J16415078-4532253 as its likely NIR counterpart.
The NIR photometry of this counterpart was performed at ESO/NTT and
is reported in CHA08. We also found the MIR counterpart in the
GLIMPSE archives (G339.1889+004889).
\newline

We observed IGR~J16418-4532 with VISIR on 2006 June 29 but did not
detect it in any filter. Using The NIR magnitudes given in CHA08, as
well as the GLIMPSE fluxes, we fitted its SED, and the
best-fitting parameters are $\textrm{A}_\textrm{v}$=14.5,
$\textrm{T}_*$=32800~K,
$\frac{\textrm{R}_\ast}{\textrm{D}_{\ast}}$=$3.77\times10^{-{11}}$,
and the reduced $\chi^2$ is 1.4/4. The best fit with the
additional component gives a larger reduced $\chi^2$ of 3.9/2
and $\textrm{T}_{\rm D}\,<\,200$ K. 

Uncertainties on the data are high, which is
the reason why the reduced $\chi^2$ are
low. Nevertheless, parameters of the fit, as well as
the 90\%-confidence ranges of parameters listed in Table 6, are
consistent with an O/B massive star nature. The temperature
of the additional component being insignificant, we
conclude this source is an O/B massive star whose enshrouding
material marginally contributes to its MIR emission.

\subsection{IGR~J16465-4507}

IGR~J16465-4507 is a transient source discovered with
\textit{INTEGRAL} on 2004
September 6-7 \citep{2004Lutovinov}. Observations were carried
out on 2004 September 14 with \textit{XMM-Newton}, and
\citet*{2004Zurita} report a
position with 4$\arcsec$ accuracy, allowing identification of an
NIR counterpart in the 2MASS catalogue
(2MASS~J16463526-4507045=USNO-B1.0~0448-00520455). With the ESO/NTT,
\citet{2005Negueruela} performed intermediate-resolution spectroscopy of
the source, estimate the spectral type is a B0.5I, and
propose that it is an SFXT. Using \textit{XMM-Newton} and
\textit{INTEGRAL}, \citet{2006Walter} find a pulse period of 227$\pm$5~s
and derive $N_{\rm H}\sim6\times10^{23}$~cm$^{-2}$.
We found its MIR counterpart in the GLIMPSE archives
(G340.0536+00.1350) using the 2MASS position.
\newline

We observed IGR~J16465-4507 with VISIR twice: 
\begin{itemize}
\item The first one on 2005
	June 20 during 600~s in PAH1. Typical seeing and airmass were
	0\farcs81 and 1.14. We 
	detected the source and the derived flux is 8.7$\pm$1.8~mJy.
\item the second on 2006 June 30 during 1200~s in PAH1 and
	PAH2. Typical seeing
	and airmass were 0\farcs68 and 1.09. We 
	detected the source in PAH1 but not in PAH2. The derived flux is
6.9$\pm$1.1
	mJy.
	
\end{itemize}
These observations show that the IGR~J16465-4507 MIR flux was constant
during the year.

Using the USNO-B1.0, 2MASS, and GLIMPSE flux values, as well as our
VISIR data, we fitted its SED,
and the best-fitting parameters are $\textrm{A}_\textrm{v}$=5.9,
$\textrm{T}_*$=25000~K,
$\frac{\textrm{R}_\ast}{\textrm{D}_{\ast}}$=$6.4\times10^{-{11}}$,
and the reduced $\chi^2$ is
13.9/7. The best fit with the additional components gives a larger
reduced $\chi^2$ of 20/5
and $\textrm{T}_{\rm D}\,<\,200$ K. 

We then conclude that no additional component is needed to explain
the MIR emission of this source, and the parameters derived
from our fit are in good agreement with IGR~J16465-4507 being a
B0.5I as reported in \citet{2005Negueruela}.

\subsection{IGR~J16479-4514}

IGR~J16479-4514 was discovered with
\textit{INTEGRAL} on 2003 August 8-9 \citep{2003Molkov}. \citet{
2005Sguera}
suggest it is a fast transient after they detected recurrent
outbursts, and \citet{2006Sguera} report a peak-flux of
$\sim\,$120~mCrab (20-60~keV). \citet{2006Walter} observed it
with \textit{XMM-Newton} and gave
its position with 4$\arcsec$ accuracy. Moreover, they derive
$N_{\rm H}\sim7.7\times10^{22}$~cm$^{-2}$ from
their observations. They also
propose 2MASS~J16480656-4512068=USNO-B1.0~0447-0531332 as its
likely NIR counterpart. The NIR spectroscopy and photometry of
this counterpart were performed 
at ESO/NTT and are reported in CHA08. It is shown that its
spectrum is consistent with an O/B supergiant. We also
found the MIR counterpart in the GLIMPSE archive
(G339.1889+004889).
\newline

We observed IGR~J16479-4514 with VISIR on 2006 June 29 in PAH1 and
PAH2, and the exposure
time was 1200~s in each filter. Typical seeing and
airmass were 0\farcs9 and 1.14. We detected it in both filters, and
the respective fluxes are 10.9$\pm$1.2~mJy and 7.0$\pm$1.6~mJy.
Using the NIR magnitudes given in CHA08, as well as the GLIMPSE
and the VISIR fluxes, we fitted its SED,
and the best-fitting parameters are $\textrm{A}_\textrm{v}$=18.5,
$\textrm{T}_*$=32800~K,
$\frac{\textrm{R}_\ast}{\textrm{D}_{\ast}}$=$1.00\times10^{-{10}}$,
and the reduced $\chi^2$ is 7.4/6. The best fit with the additional
component gives a larger
reduced $\chi^2$ of 9/4
and $\textrm{T}_{\rm D}\,<\,200$ K.

We then do not need any additional component to fit the SED, and
our result is consistent with IGR~J16479-4514 being an
obscured O/B supergiant, in good agreement with CHA08. 

\subsection{IGR~J17252-3616}

IGR~J17252-3616 is a heavily-absorbed persistent source
discovered with \textit{INTEGRAL} on 2004 February 9 and
reported in \citet{2004Walter}. It was observed with
\textit{XMM-Newton} on 2004 March 21, and \citet{2006Zurita} give its
position with 4$\arcsec$ accuracy. Using the
\textit{XMM-Newton} observations, as well
as those carried out with \textit{INTEGRAL}, they show the source was a
binary X-ray pulsar with a spin period of $\sim\,$413.7~s and an orbital
period of $\sim\,9.72$ days, and derive
$N_{\rm H}\sim1.5\times10^{23}$~cm$^{-2}$. Moreover, they
fitted its high-energy spectrum with either an absorbed compton
(kT$\sim$ 5.5~keV and $\tau\sim7.8$) or a flat power law
($\Gamma\sim0.02$). 

In their paper, they
propose 2MASS~J17251139-3616575 as its 
likely NIR counterpart, as do \citet*{2007Negueruela}. The NIR
spectroscopy and photometry of
this counterpart were performed 
at ESO/NTT and are reported in CHA08, where it is shown that its
spectrum is consistent with an O/B supergiant. Using the
2MASS position, we searched for its MIR counterpart in the GLIMPSE
catalogue. Unfortunately, we did not find its IRAC fluxes
in the database. Nevertheless, we found post-Basic Calibrated
Data (post-BCD) images of the source in all
filters. We then reduced those data and derived fluxes directly from
the images. They are listed in Table 3.
\newline

We observed IGR~J17252-3616 with VISIR on 2006 June 30 in PAH1 and
PAH2 and the exposure
time was 1200~s in each filter. Typical seeing and
airmass were 0\farcs97 and 1.09. We detected it in PAH1, and the
derived
flux is 6.1$\pm$0.6~mJy. Using the NIR magnitudes given in CHA08,
as well as the GLIMPSE and the VISIR fluxes, we fitted its SED,
and the best-fitting parameters are $\textrm{A}_\textrm{v}$=20.8,
$\textrm{T}_*$=32600~K,
$\frac{\textrm{R}_\ast}{\textrm{D}_{\ast}}$=$7.57\times10^{-{11}}$,
and the reduced $\chi^2$ is 3.8/5. The best fit with the additional
component gives a larger
reduced $\chi^2$ of 6.9/3
and $\textrm{T}_{\rm D}\,<\,200$ K. We then do not
need any additional component to fit the SED, and
our result is consistent with IGR~J17252-3616 to be an
obscured O/B supergiant, in good agreement with CHA08. 

\subsection{IGR~J17391-3021}

IGR~J17391-3021 is a transient source discovered with
\textit{INTEGRAL} on 2003
August 26 \citep{2003Sunyaev} and it corresponds to the \textit{Rossi
X-ray Timing Explorer} (\textit{RXTE}) source
XTE J1739-302. \citet{2005Sguera} analysed archival
\textit{INTEGRAL} data and
classified the source as a fast X-ray transient presenting a typical
neutron star spectrum. \citet{2006Smith} observed it with
\textit{Chandra} on
2003 October 15 and give its precise position with
1$\arcsec$ accuracy.
They also give its optical/NIR counterpart
2MASS~J17391155-3020380=USNO-B1.0~0596-0585865 and classify IGR
J17391-3021 as an SFXT. \citet{2006Negueruelab}
performed optical/NIR photometry and spectroscopy of the companion
using ESO/NTT and find it is a O8Iab(f) star whose distance is
$\sim$ 2.3 kpc. CHA08 also report optical and NIR spectroscopy
and photometry of the companion carried out at ESO/NTT and
confirm the nature of the companion. Using the
2MASS position, we searched for its MIR counterpart in the GLIMPSE
catalogue and as for IGR~J17252-3616, we had to reduce post-BCD data
and derive the fluxes directly from the images. The fluxes are
listed in Table 3.
\newline

We observed IGR~J17391-3021 with VISIR on 2005 June 20 in PAH1 and
PAH2, and the exposure
time was 600~s in each filter. Typical seeing and
airmass were 0\farcs63 and 1.13. We detected it in both filters, and
the derived fluxes are 70.2$\pm$1.6~mJy and 46.5$\pm$2.6~mJy. Using
the optical and the NIR magnitudes given in CHA08, as well as
the GLIMPSE and
the VISIR fluxes,
we fitted its SED and the best-fitting parameters are
$\textrm{A}_\textrm{v}$=9.2,
$\textrm{T}_*$=31400~K,
$\frac{\textrm{R}_\ast}{\textrm{D}_{\ast}}$=$1.8\times10^{-{10}}$,
and the reduced $\chi^2$ is 11.7/10. The best fit with the
additional component gives a larger
reduced $\chi^2$ of 15.3/8
and $\textrm{T}_{\rm D}\,<\,200$ K.

We then do not need any additional component to fit the SED, and the
parameters derived from our fit are in good agreement with
IGR~J17391-3021 to be an O8Iab(f) supergiant star, as initially
reported in \citet{2006Negueruelab}.

\subsection{IGR~J17544-2619}

IGR~J17544-2619 is a transient source discovered with
\textit{INTEGRAL} on 2003
September 17 \citep{2003Sunyaev}. \citet{2004Gonzalez-Riestra}
observed it with \textit{XMM-Newton} and derive
$N_{\rm H}\sim2\times10^{22}$~cm$^{-2}$. They also confirm the
association of the companion with
2MASS~J17542527-2619526=USNO-B1.0~0636-0620933, as proposed
in \citet{2003Rodriguezb}. \citet{2005Zand} report on
observations performed with \textit{Chandra}, give its position
with 0\farcs6
accuracy, $N_{\rm H}\sim$ 1.36$\times10^{22}$~cm$^{-2}$, and show
that its high-energy spectrum is typical of an accreting
neutron star. Moreover, they identify the counterpart as a blue
supergiant. \citet{2006Sguera} report a peak-flux of
$\sim\,$240~mCrab. 
Using ESO/NTT, \citet{2006Pellizza} performed
optical/NIR spectroscopy and photometry of the companion, and give
its spectral type as O9Ib at 2.1-4.2 kpc. Using the 2MASS
position, we searched for its MIR counterpart in the GLIMPSE
catalogue, and as for IGR~J17252-3616 and IGR~J17391-3021, we had
to reduce post-BCD data
and derived the fluxes directly from the images. The fluxes are listed
in Table 3.
\newline

We observed IGR~J17544-2619 with VISIR on 2005 June 20 in PAH1 and
PAH2, and the exposure
time was 600~s in PAH1 and 1200~s in PAH2. Typical seeing and
airmass were 0\farcs64 and 1.13. We detected it in both filters, and
the derived fluxes are 46.1$\pm$2.8~mJy and 20.2$\pm$2.1~mJy. Using
the magnitudes from \citet{2006Pellizza}, the GLIMPSE and the VISIR fluxes,
we fitted its SED, and the best-fitting parameters are
$\textrm{A}_\textrm{v}$=6.1,
$\textrm{T}_*$=31000~K,
$\frac{\textrm{R}_\ast}{\textrm{D}_{\ast}}$=$1.27\times10^{-{10}}$,
and the reduced $\chi^2$ is 6.1/8. The best fit with the additional
component gives a larger
reduced $\chi^2$ of 9/6 and $\textrm{T}_{\rm D}\,<\,200$ K.

We then do not need any additional component to fit the SED, and the
parameters derived from our fit are in good agreement with
IGR~J17544-2619 as an O9Ib supergiant star, as initially
reported in \citet{2006Pellizza}.

\subsection{IGR~J19140+0951}

IGR~J19140+0951 is a persistent source that was discovered with
\textit{INTEGRAL} on 2003 March 6-7
\citep{2003Hannikainen}. Observations carried out with
\textit{RXTE} allowed $\Gamma\sim1.6$
and $N_{\rm H}\sim6\times10^{22}$~cm$^{-2}$ to be derived 
\citep*{2003Swank}. Timing analysis of the \textit{RXTE} data
showed a period of 13.55 days
\citep{2004Corbet}, which shows the binary nature of the
source. After a comprehensive analysis of
\textit{INTEGRAL} and \textit{RXTE} data, \citet{2005Rodriguez}
show the source is spending most of
its time in a faint state but report high variations in luminosity
and absorption column density (up to $\sim\,10^{23}$~cm$^{-2}$). They
also find evidence that the compact object is a neutron star rather
than a black hole. Using \textit{Chandra} observations carried
out 2004 May 11,
\citet{2006Zand} give its position with 0\farcs6 accuracy.
This allowed them to find its NIR counterpart in the 2MASS
catalogue (2MASS~J19140422+0952577). Moreover, they searched for
its MIR counterpart in the Mid-course Space Experiment
\citep[MSX][]{1994Mill}
and found an object at 8.3 $\mu$m. The NIR photometry and
spectroscopy of this source were performed at
ESO/NTT and results are reported in CHA08. It is shown that its
spectrum is consistent with an O/B massive star, in good
agreement with \citet{2007Nespoli}, who show it is a B1I supergiant.
Using the 2MASS position, we also found its MIR
counterpart in the GLIMPSE archive (G044.2963-00.4688).
\newline

We observed IGR~J19140+0951 with VISIR on 2006 June 30 in PAH1 and
PAH2, and the exposure
time was 1200~s in each filter. Typical seeing and
airmass were 1\farcs12 and 1.17. We detected it in both filters,
and the derived
fluxes are 35.2$\pm$1.4~mJy and 19.1$\pm$1.4~mJy. We point out that
the object given as the MSX counterpart of IGR~J19140+0951 in
\citet{2006Zand} is a blended source. Indeed, VISIR
images, whose resolution is far better, clearly show there are
two sources in the field,
IGR~J19140+0951 and a very bright southern source (see Fig 2). 

\begin{figure}
	\centering
	\includegraphics[width=5cm]{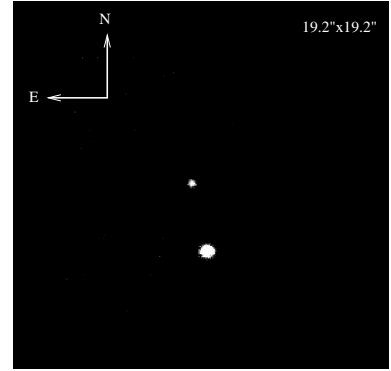}
	\caption{VISIR image of IGR~J19140+0951 in PAH1 (8.59
	$\mu$m). 19\farcs2x19\farcs2 field of view and 0\farcs075
	plate scale. We clearly see the two sources that were blended
	with MSX. The MIR counterpart of IGR~J19140+0951 is the
	northern source.}
\end{figure}

Using the magnitudes
given in CHA08, as well as the fluxes from GLIMPSE and our
observations with VISIR, we fitted its SED,
and the best-fitting parameters are $\textrm{A}_\textrm{v}$=16.5,
$\textrm{T}_*$=22500~K,
$\frac{\textrm{R}_\ast}{\textrm{D}_{\ast}}$=$1.92\times10^{-{10}}$, and
the reduced $\chi^2$ is 14.4/6. The best fit with the additional
component gives a larger
reduced $\chi^2$ of 20.2/4 and $\textrm{T}_{\rm D}\,<\,200$ K.

We do not need any additional component to fit the SED, and the
parameters derived from our fit are in good agreement with
IGR~J19140+0951 to be an B1I supergiant star, as initially
reported in \citet{2007Nespoli}.

\begin{figure*}
	\centering
	\begin{tabular}{c c c}
	\includegraphics[angle=-90,width=6.cm]{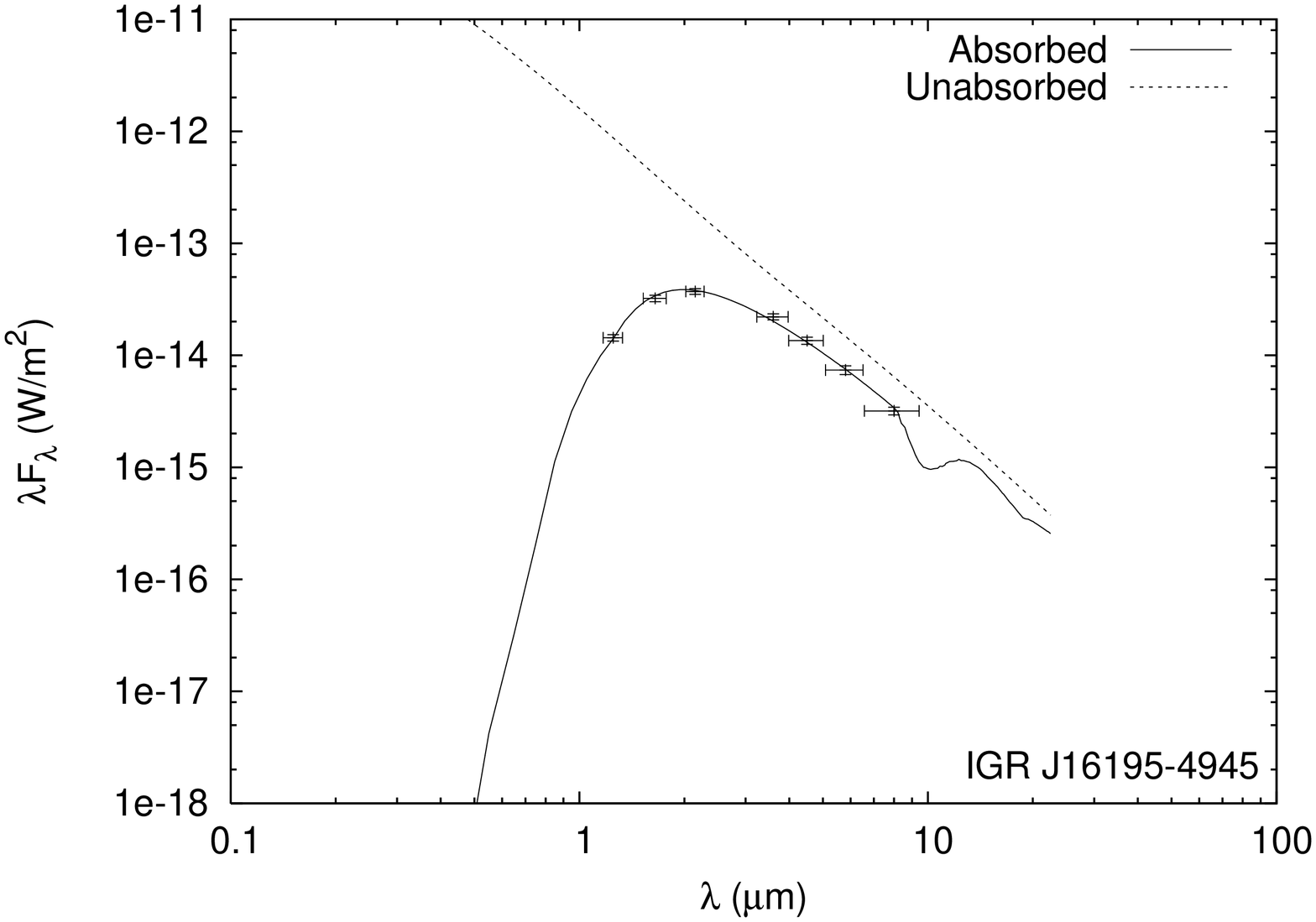}&
	\includegraphics[angle=-90,width=6.cm]{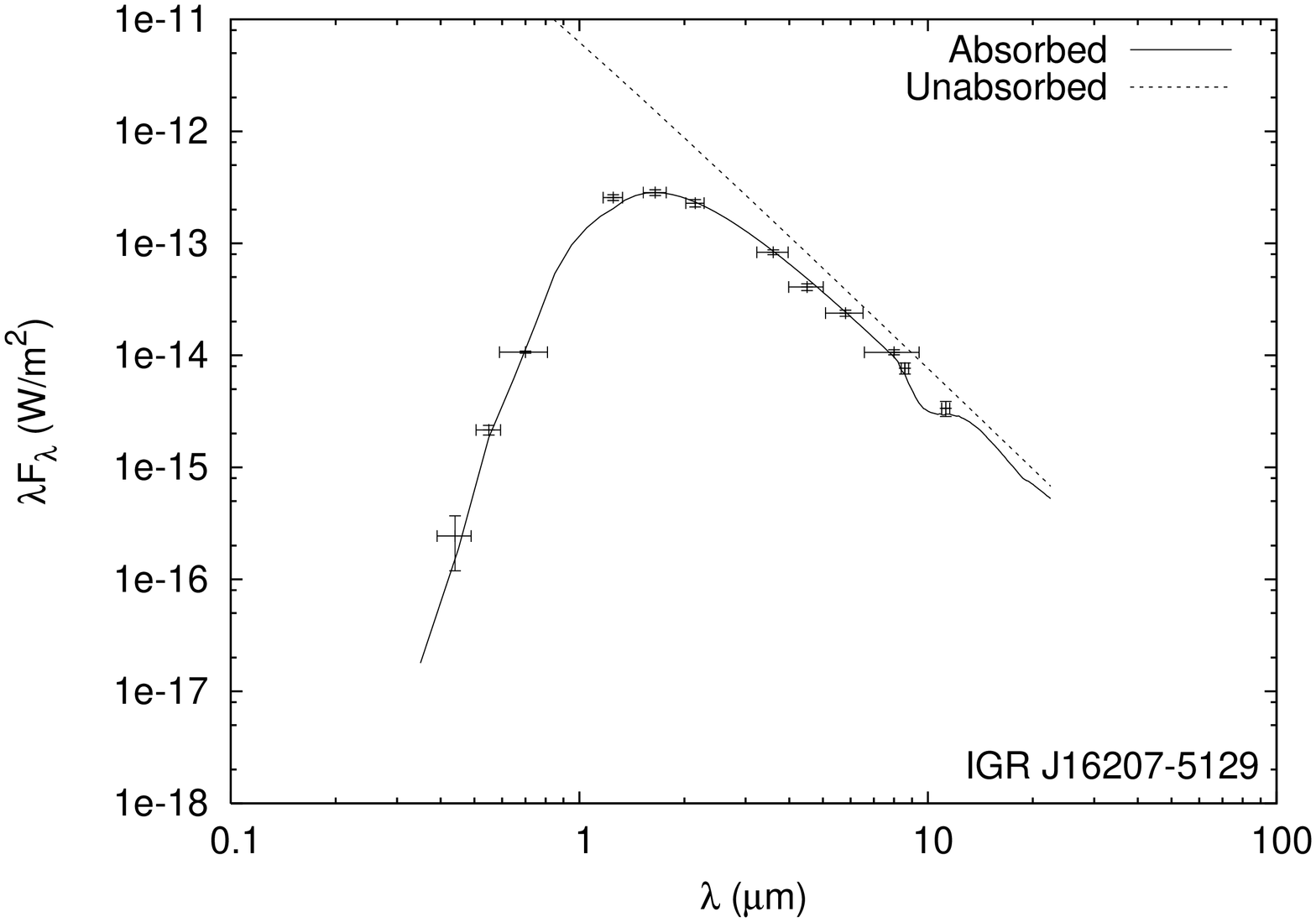}&
	\includegraphics[angle=-90,width=6.cm]{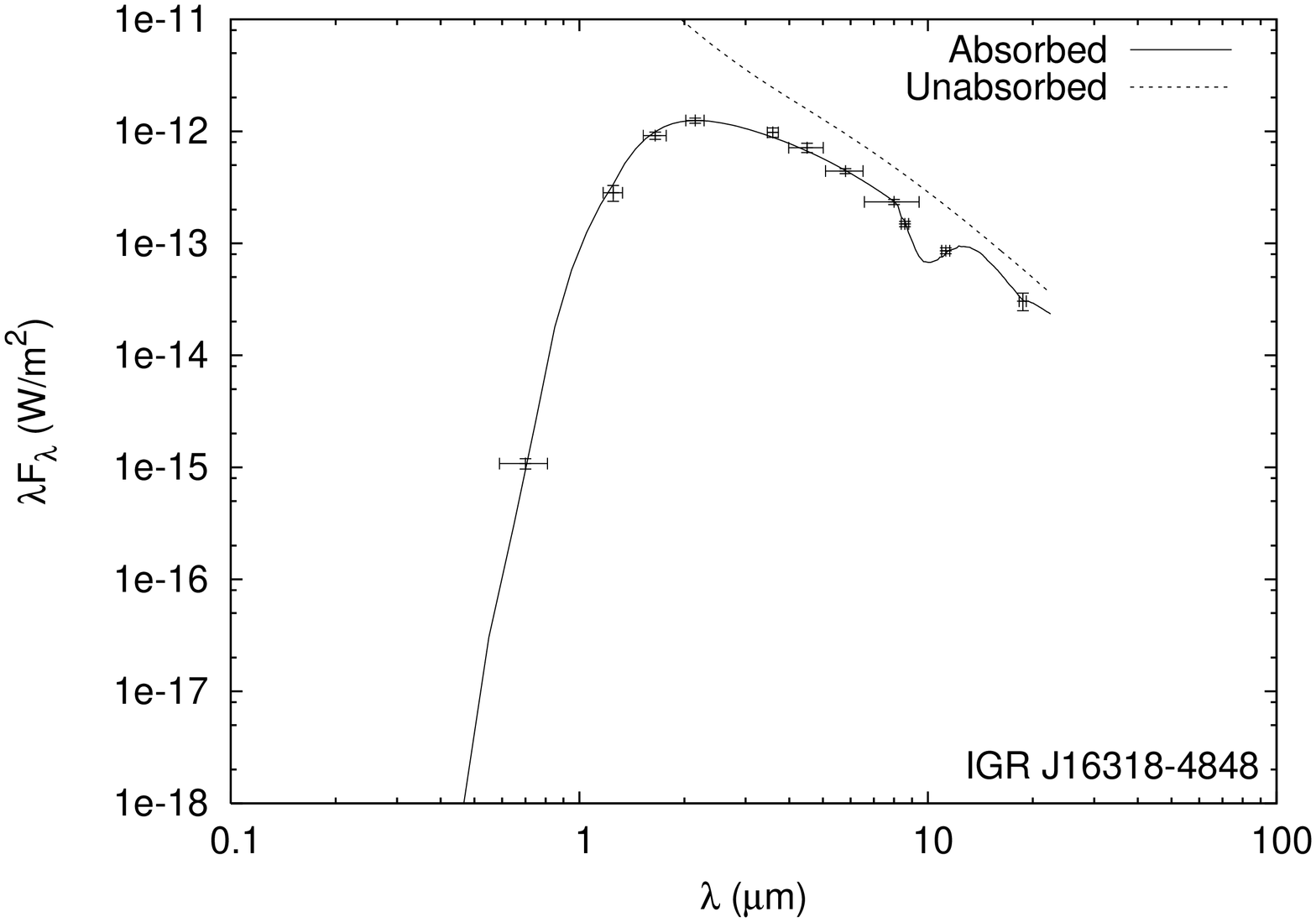}\\
	\includegraphics[angle=-90,width=6.cm]{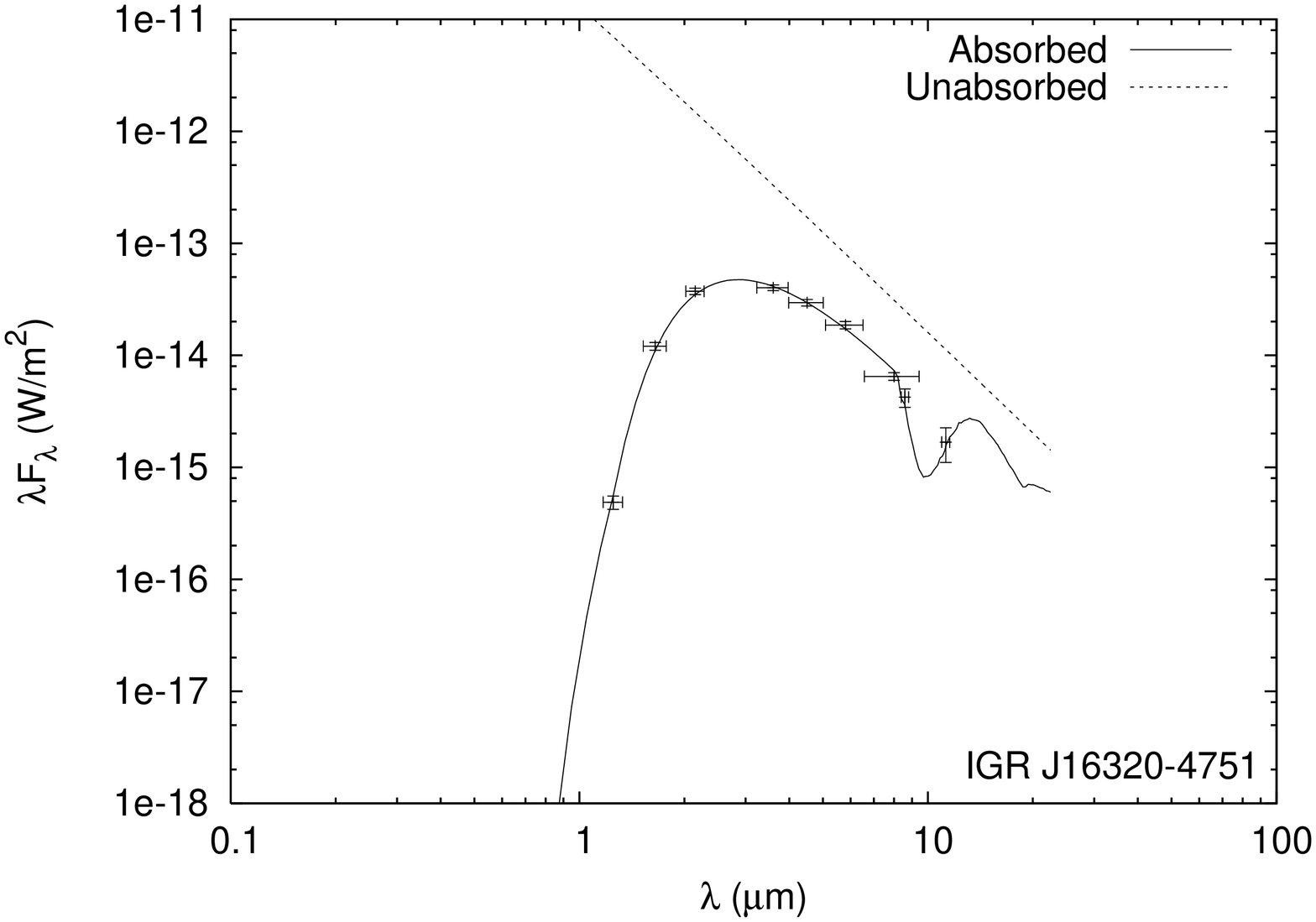}&
	\includegraphics[angle=-90,width=6.cm]{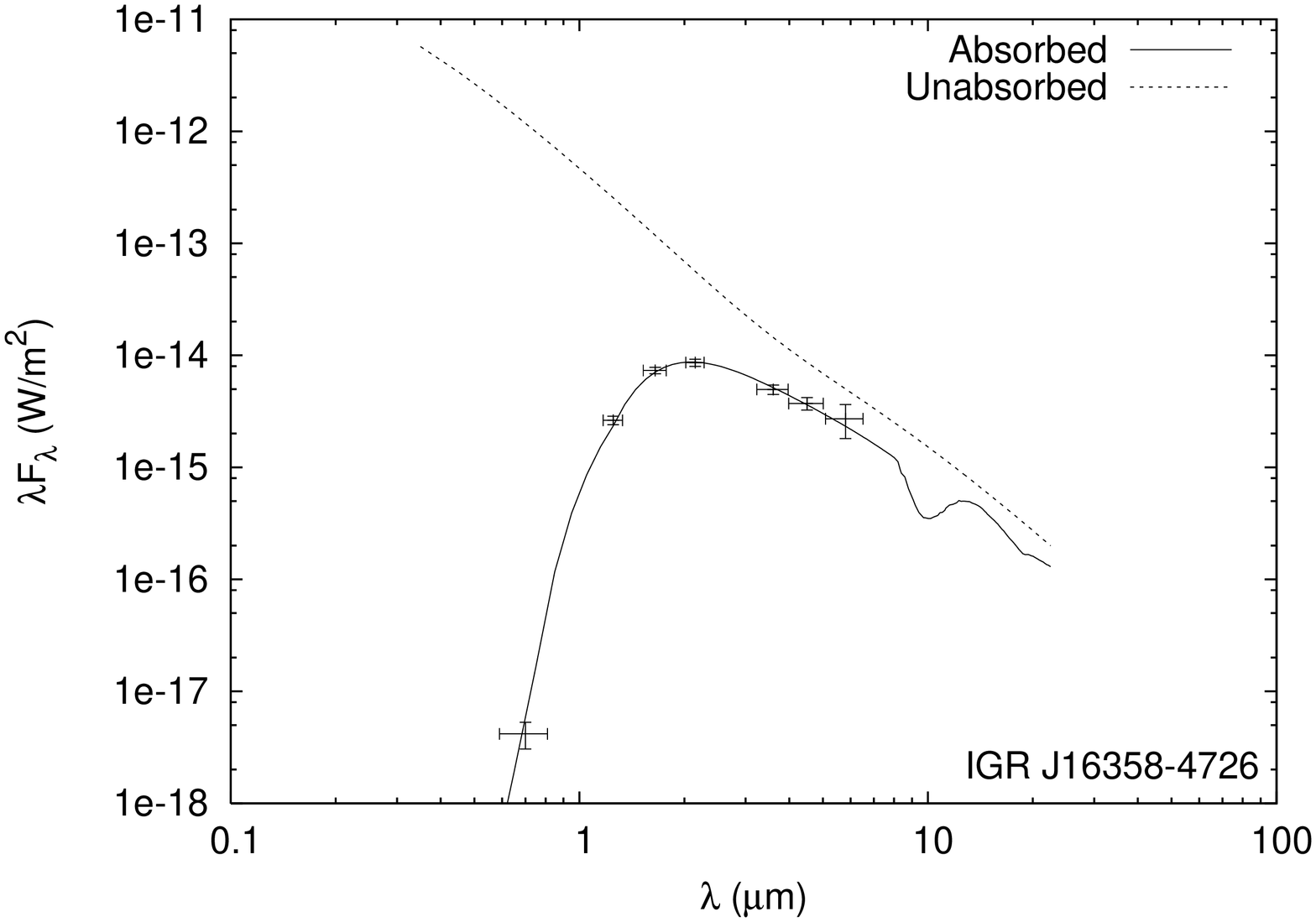}&
	\includegraphics[angle=-90,width=6.cm]{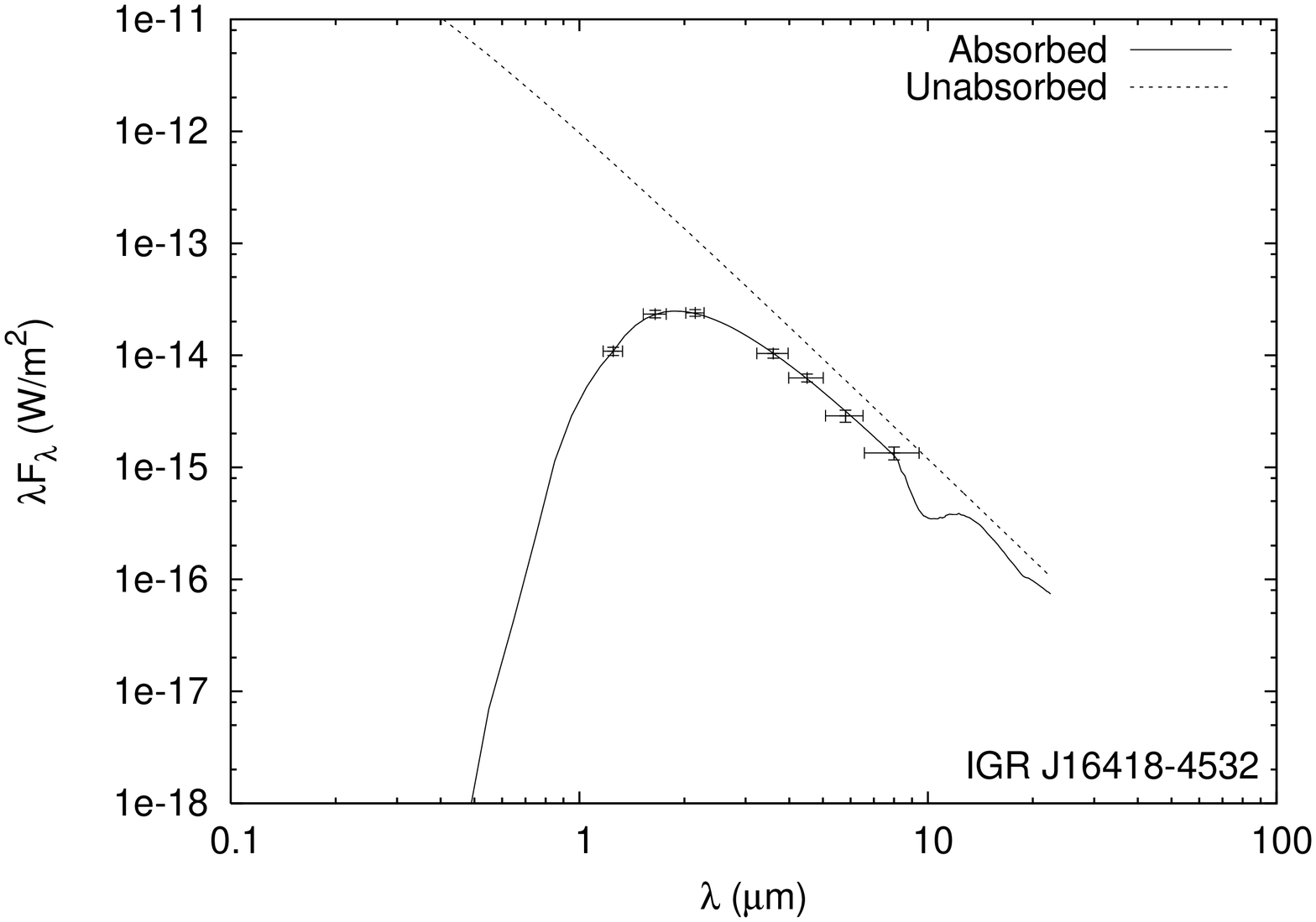}\\
	\includegraphics[angle=-90,width=6.cm]{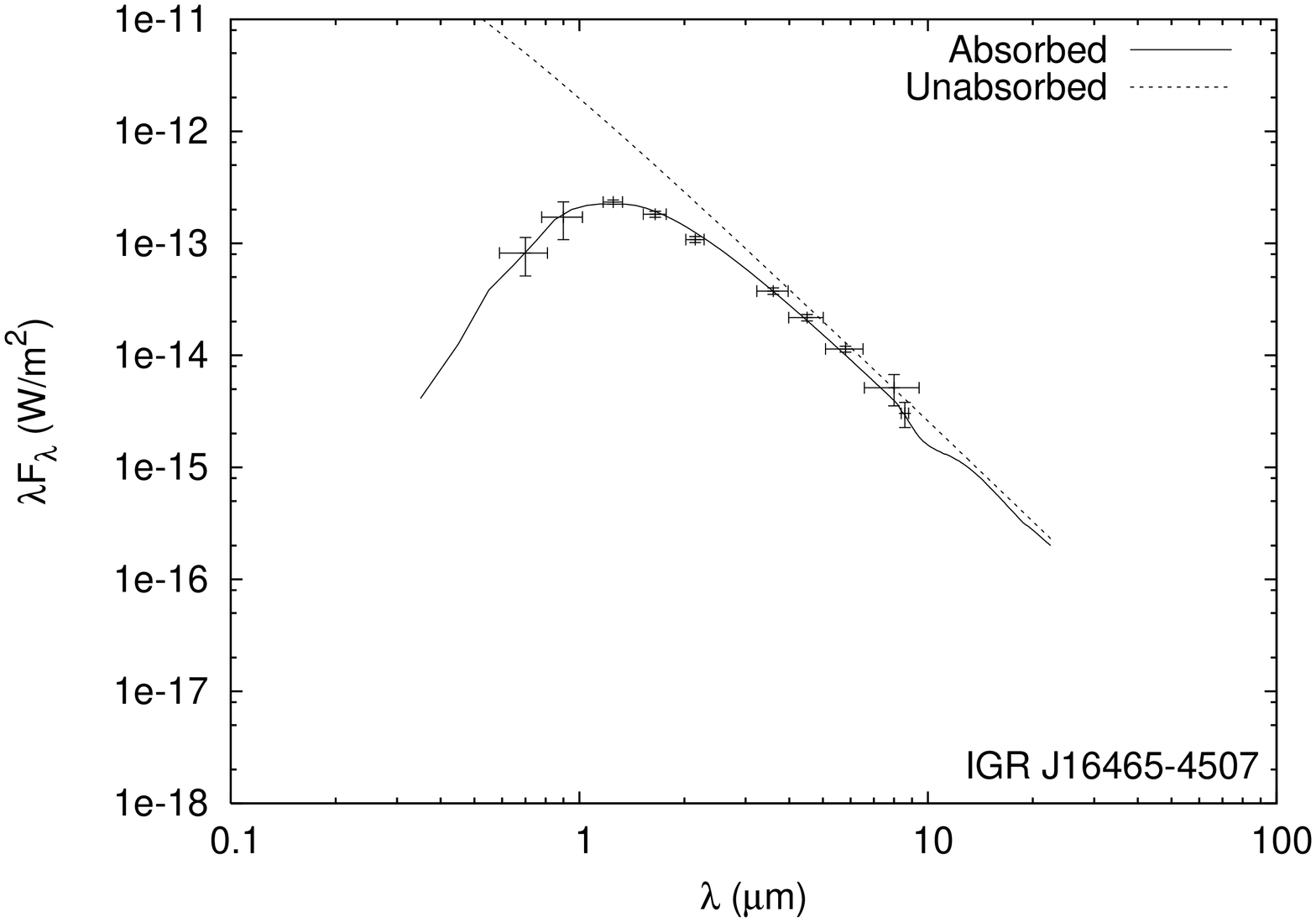}&
	\includegraphics[angle=-90,width=6.cm]{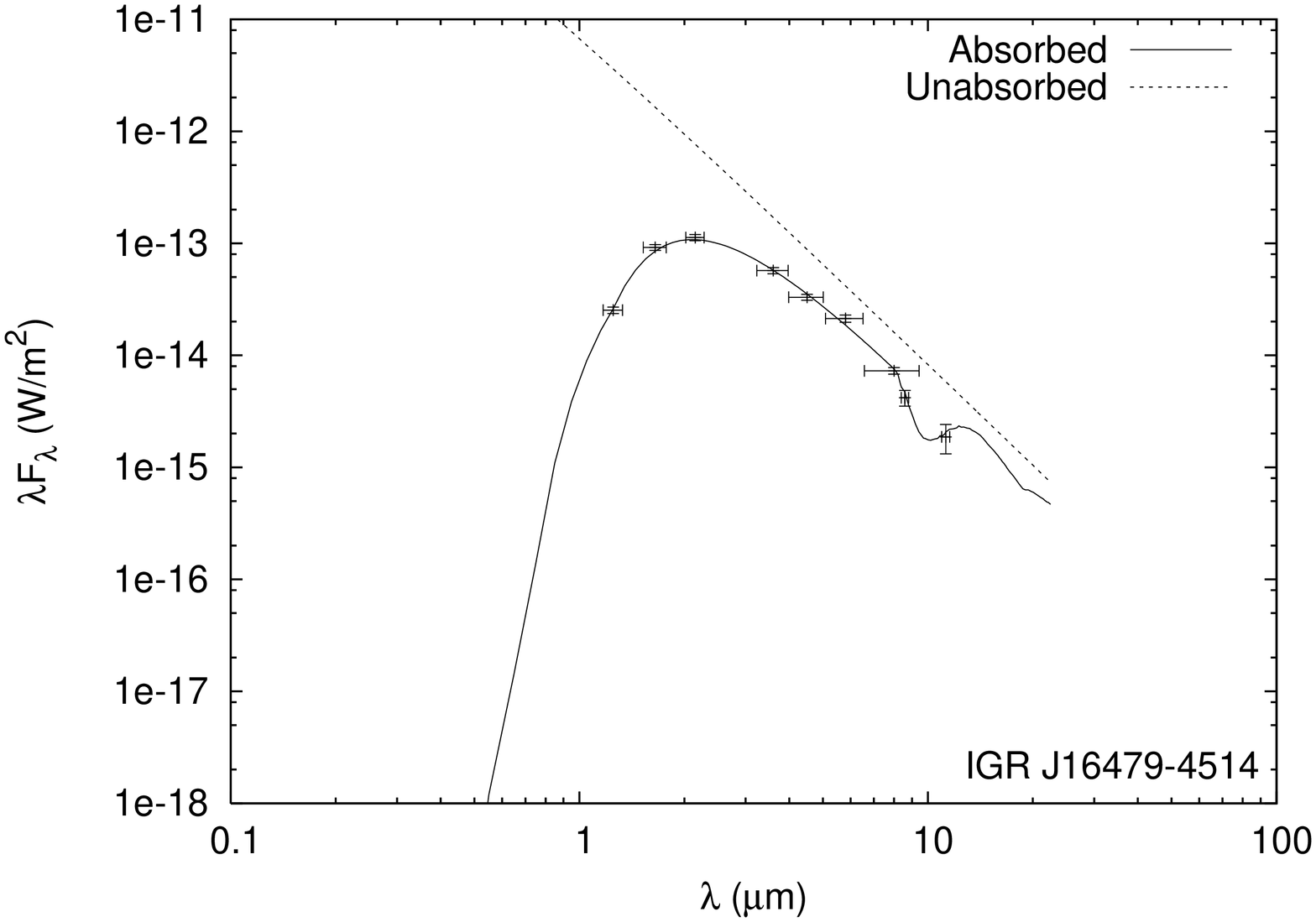}&
	\includegraphics[angle=-90,width=6.cm]{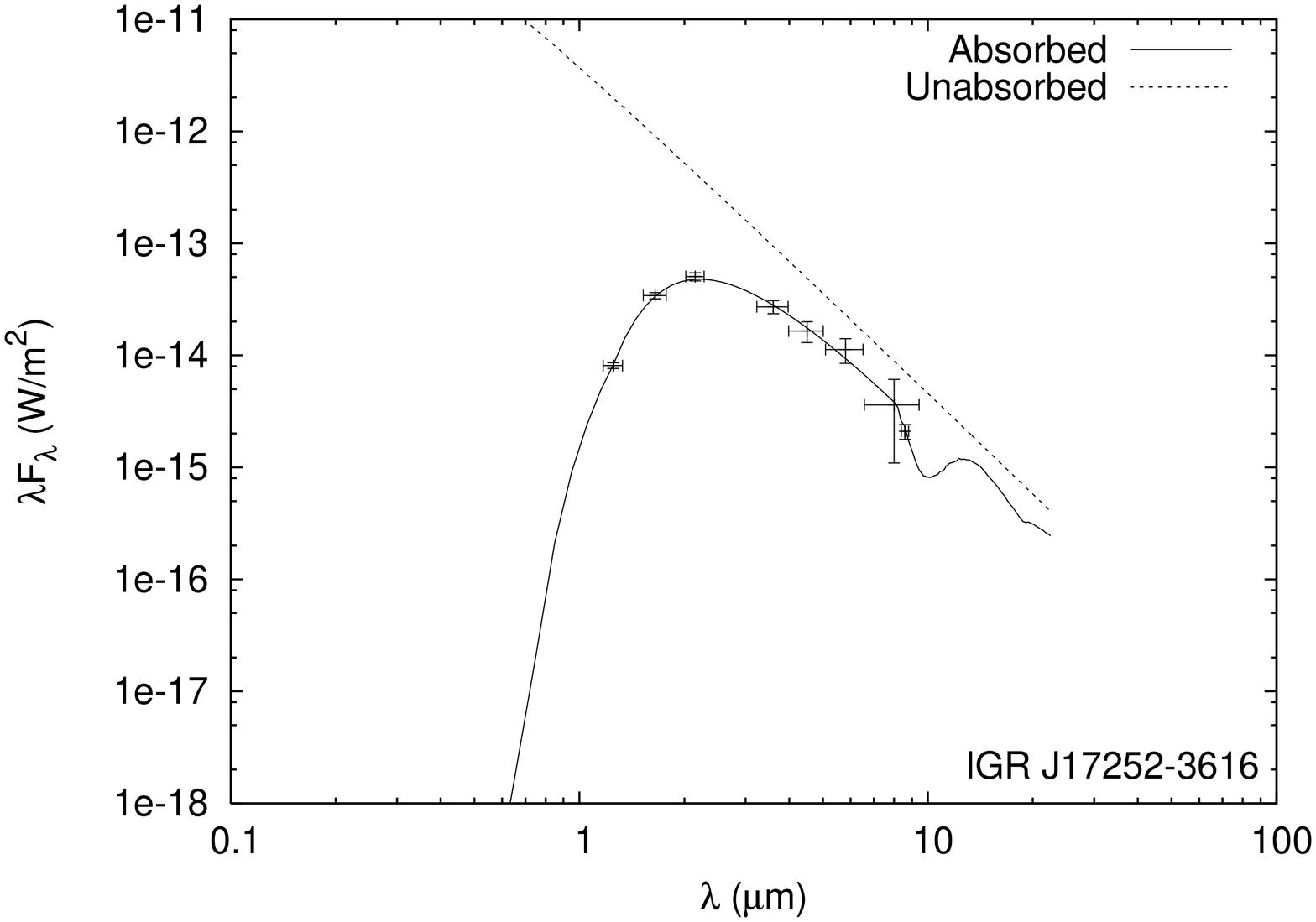}\\
	\includegraphics[angle=-90,width=6.cm]{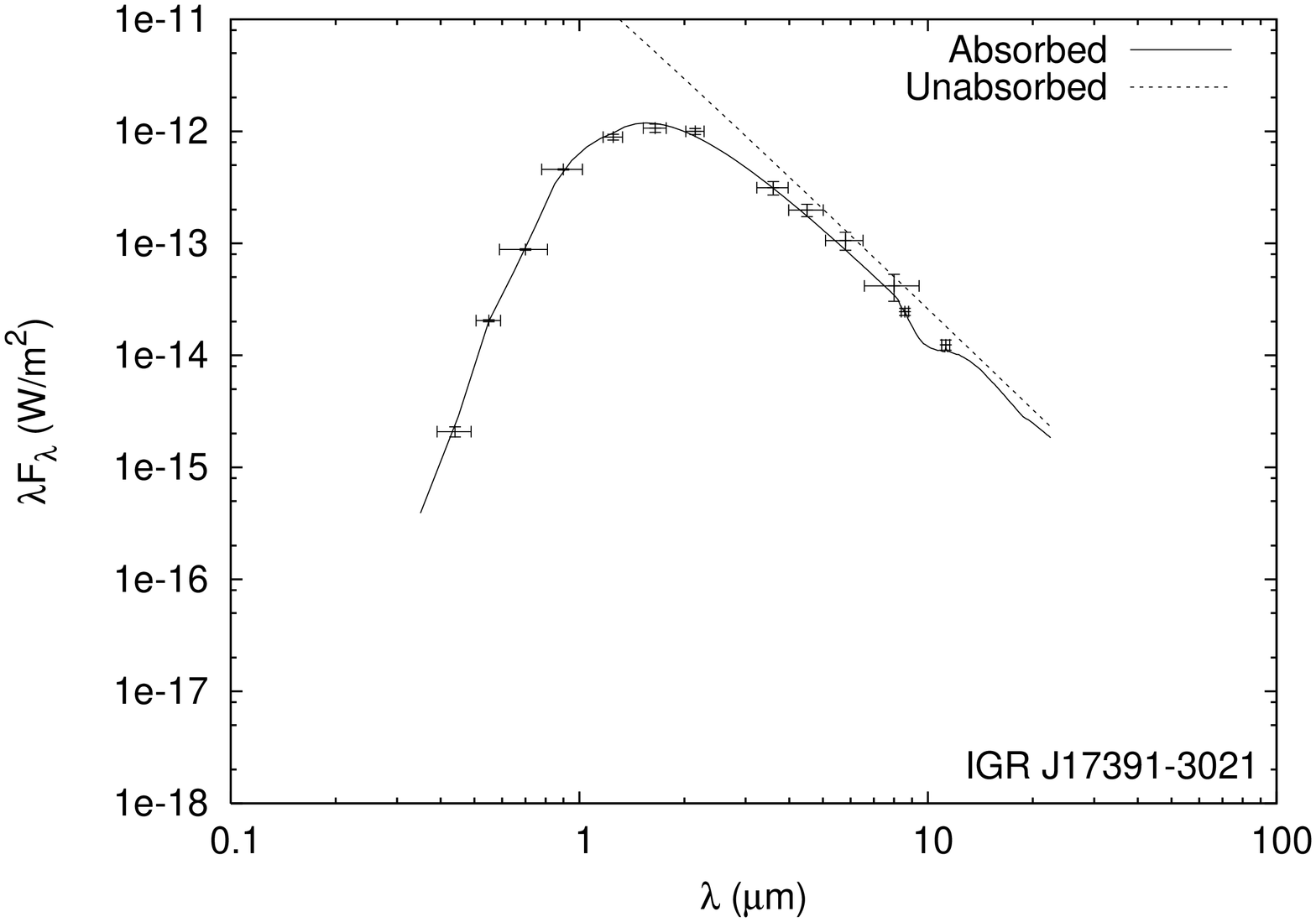}&
	\includegraphics[angle=-90,width=6.cm]{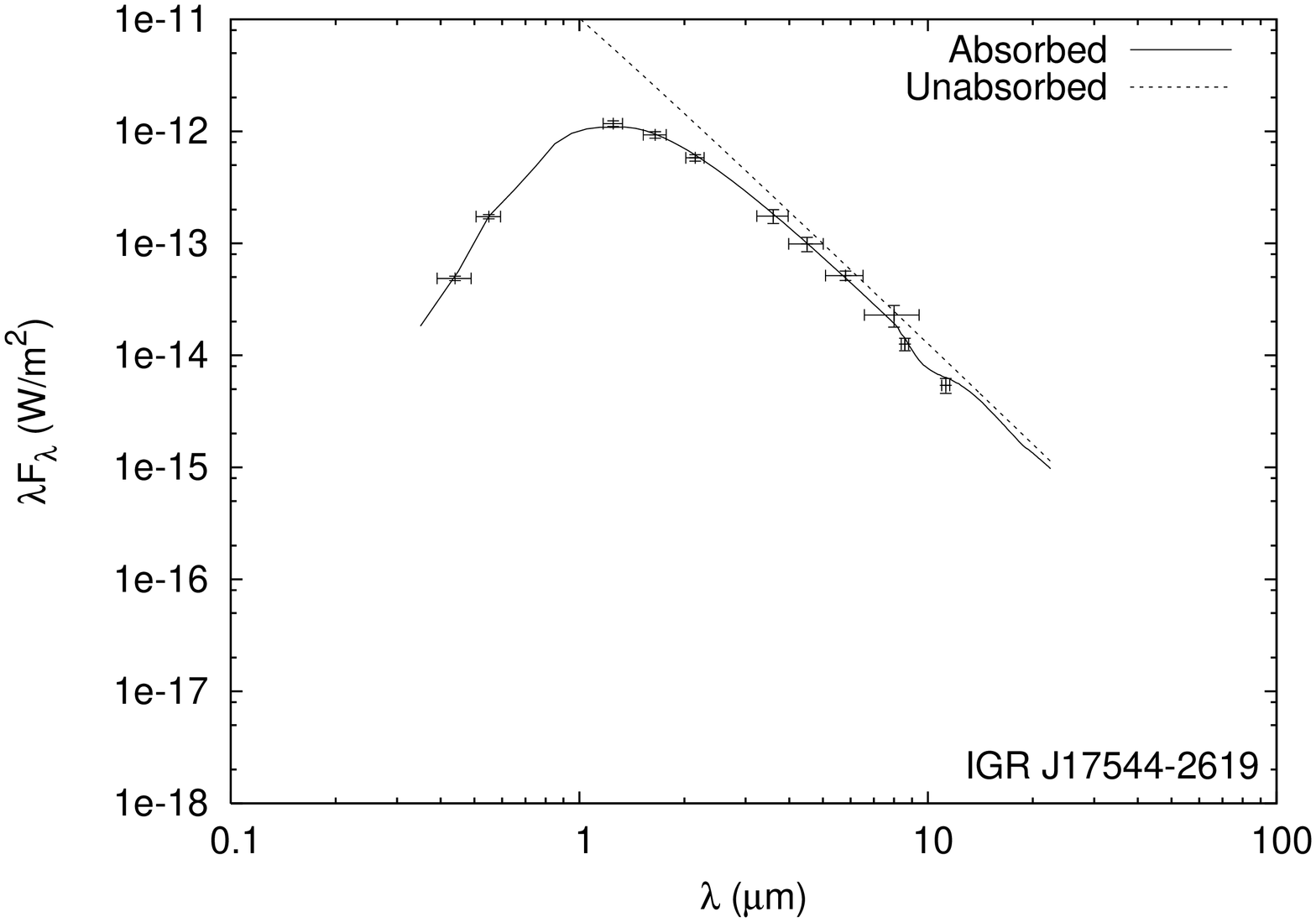}&
	\includegraphics[angle=-90,width=6.cm]{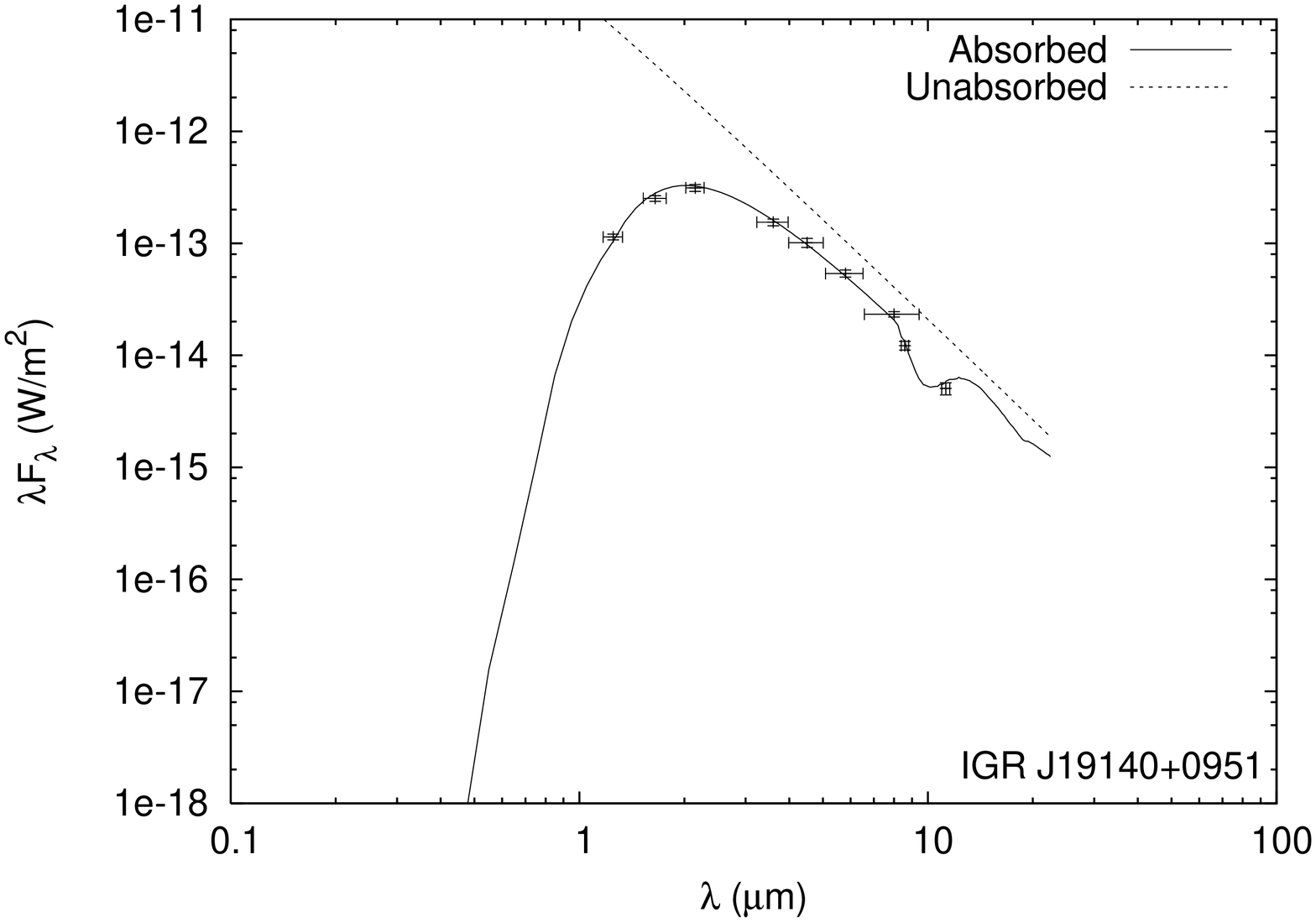}\\
	\end{tabular}
	\caption{\small Optical-to-MIR absorbed (line) and unabsorbed
	(dotted-line) SEDs of 12 \textit{INTEGRAL} sources,
including broad-band photometric data
	from ESO/NTT, 2MASS, GLIMPSE, and VISIR.}
\end{figure*}

\begin{figure}
	\centering
	\includegraphics[angle=-90,width=8.5cm]{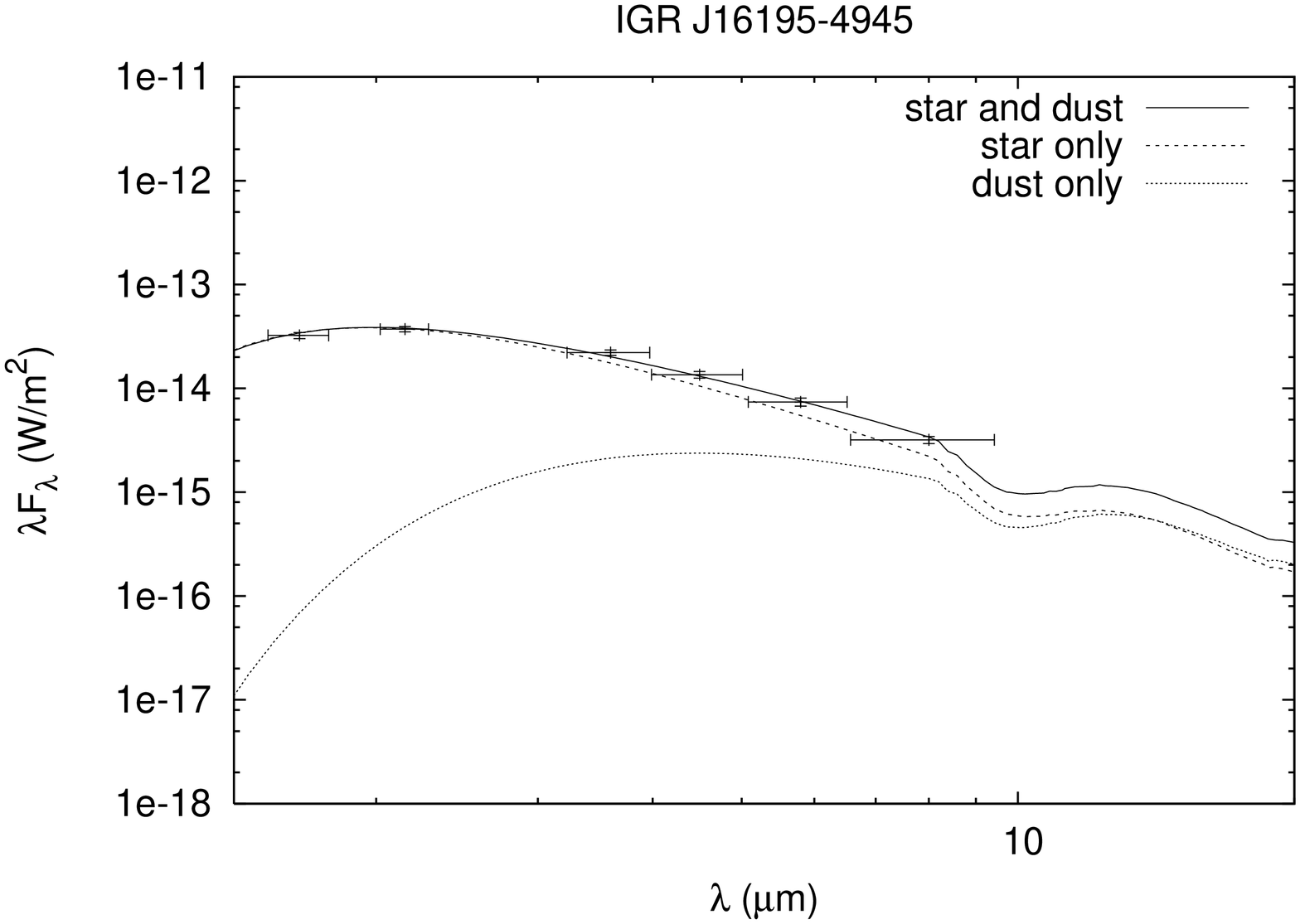}
	\includegraphics[angle=-90,width=8.5cm]{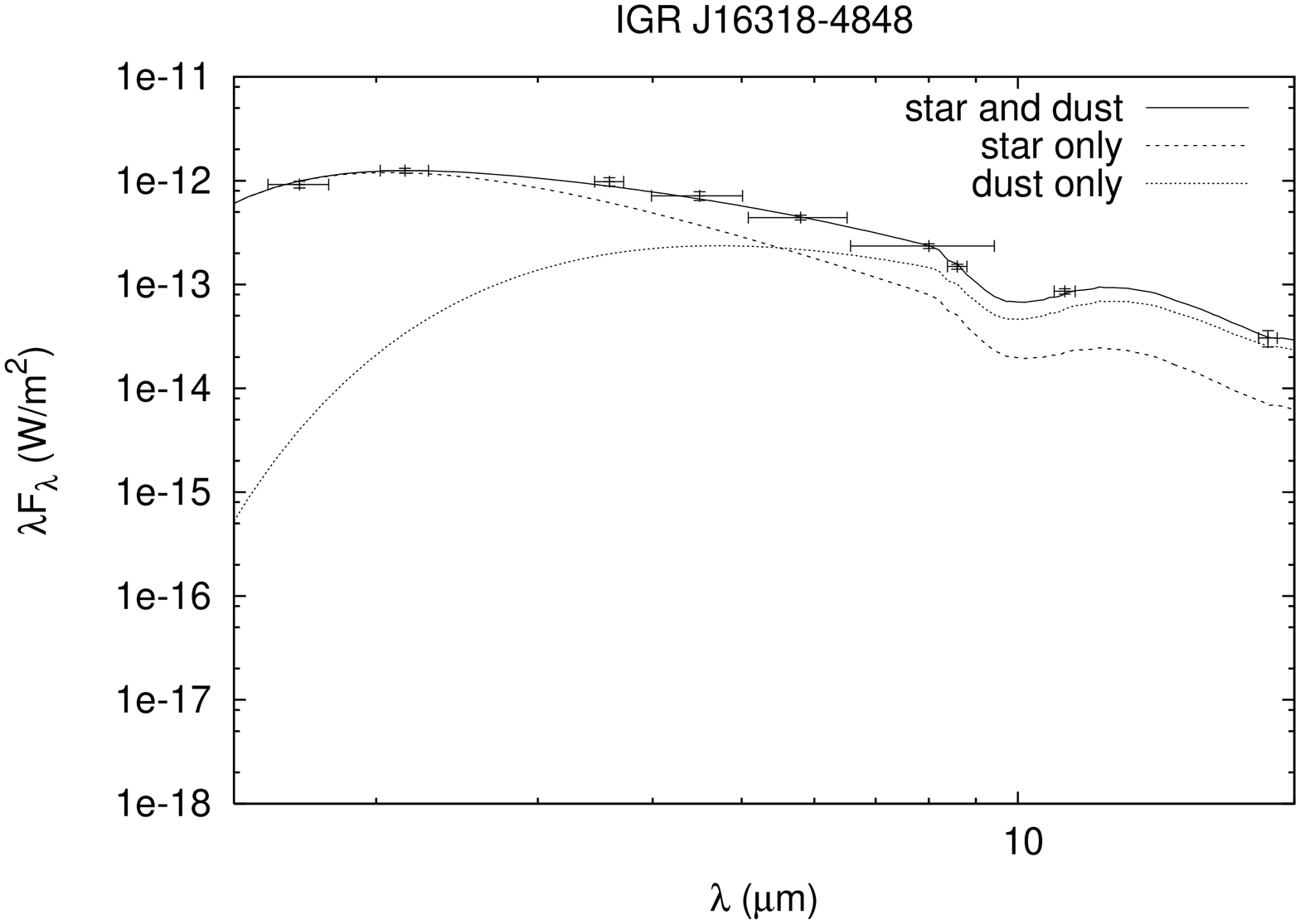}
	\includegraphics[angle=-90,width=8.5cm]{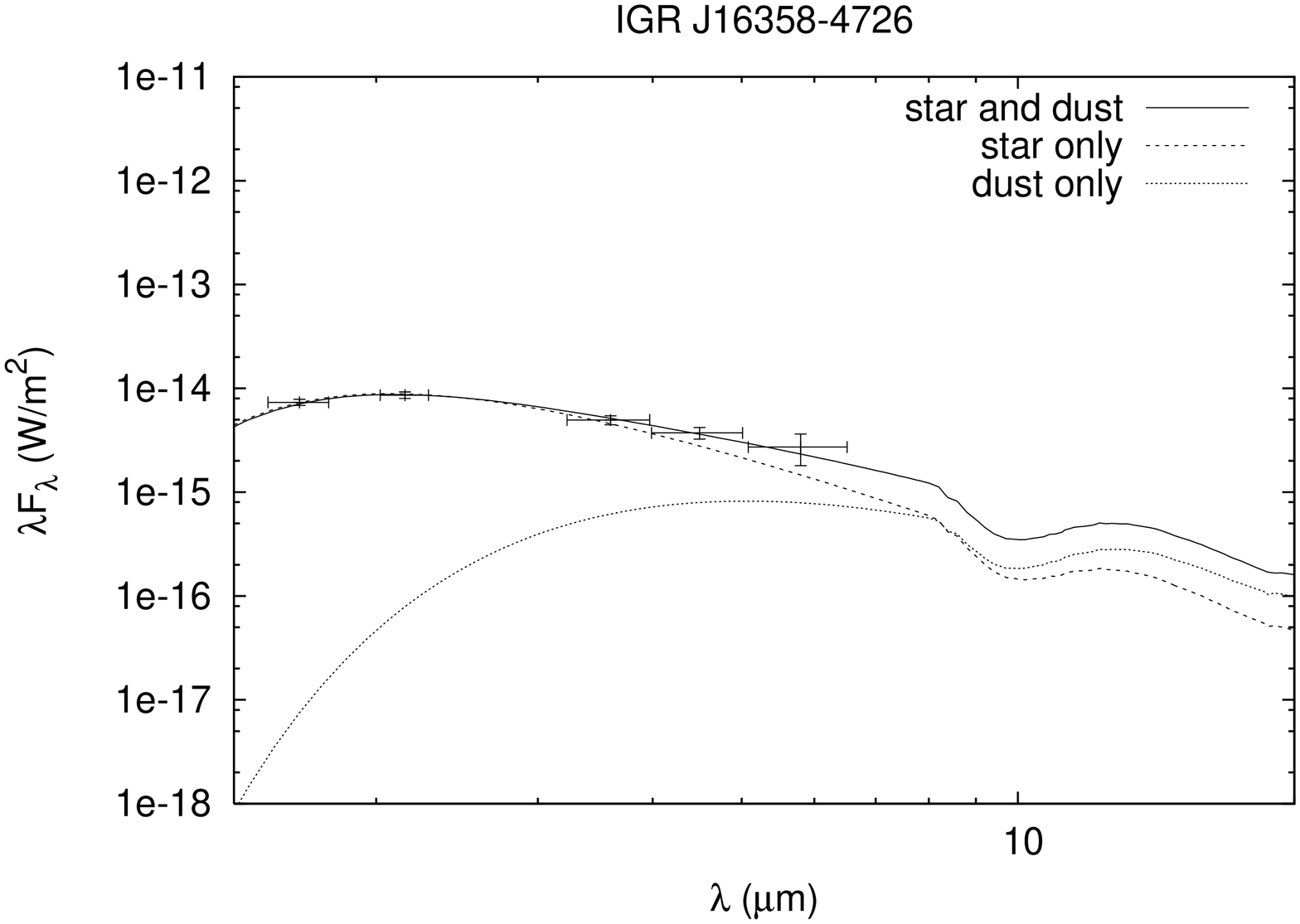}
	\caption{IR SEDs of IGR~J16195-4945,
	IGR~J16318-4848, and IGR~J16358-4726 in the NIR and the MIR. We
	show their SEDs including the contribution of the star and the
	dust (line), the star only (dashed-line), and the dust only
(dotted-line).}
\end{figure} 

\section{Discussion}

\subsection{B vs B[e] supergiants stellar winds}

All SEDs were best-fitted without any dust component (even
the very absorbed one like IGR~J16320-4751), except three of them
(IGR~J16195-4945, IGR~J16318-4848, and
IGR~J16358-4726, see Fig. 3) that exhibit a MIR excess likely
due to the presence of dust in their stellar wind. 

Blue supergiants are known to
exhibit a very strong but sparse stellar wind of high
velocity ($\sim$ 1000-2000~km~s$^{-1}$). This has been
explained through the so-called
radiation line-driven CAK model \citep*{1975Castor} in which the
wind is driven by
absorption in spectral lines. Hot stars emit most of
their radiation in the ultraviolet (UV) where their atmosphere
has many absorption lines. Photons coming from the
photosphere of the star with the same wavelength are absorbed and
re-emitted to the expanding medium in a random direction with
almost the same
momentum, which results in acceleration of the wind. This process
is very effective because the line
spectrum of the scattering ions in the wind is Doppler-shifted
compared to the stellar rest frame, so the scattering atoms are
shifted with respect to their neighbours at lower velocities and
can interact with an unaffected part of the stellar spectrum.
\newline

IGR~J16318-4848 was proven to belong to a particular class of B1
supergiants, the B[e] supergiants or sgB[e]
\citep*{2004Filliatre}. 
A physical definition of B[e] stars can be found in
\citet{1998Lamers}. We just recall two of the
characteristics here: the presence of forbidden emission lines
of [\ion{Fe}{ii}] and [\ion{O}{i}] in the NIR spectrum and 
of a strong MIR excess due to hot circumstellar dust 
that re-emits the absorbed stellar radiation through
free-free emission.
An sgB[e] is defined by the B[e] phenomenon, the
indication of mass-loss in the optical spectrum (P-cygni
profiles), and a hybrid spectrum characterised by the
simultaneous presence of narrow low-excitation lines and broad
absorption features of high-excitation lines. This hybrid nature
was empirically explained by the simultaneous presence of a
normal supergiant hot polar wind (fast and sparse) and
responsible for the broad lines and a cool equatorial
outflowing disk-like
wind (slow and dense) responsible for the narrow lines
\citep{1983Shore,1985Zickgraf,1987Shore}. This
empirical model has received some confirmation from polarimetry
\citep*{1999Oudjmaier}.
\newline

There are a few models that explain the creation of this outflowing
disk, and all of them consider the
star rotation to be an important parameter in the process. In
this paper, we present only the most consistent of them,
the Rotation
Induced Bi-stability mechanism (RIB), but a review can be found
in \citet*{2006Kraus}. 
\newline

The lines responsible for the creation of the wind are dependent
on the ionisation structure, and a change in this structure leads to
a change in the radiative flux. This is the bi-stability jump
found by \citet*{1991Lamers}, which appears
for B stars with effective temperatures
of about 23000~K. Above this temperature, the wind tends to be fast
and sparse. Below, the mass-loss rate if five times higher and
the terminal velocity two
times slower, which leads to a wind that is ten times denser. 

\citet*{1997Cassinelli} propose that
the same effect is important from polar to equatorial regions
for rapidly
rotating B stars. Indeed, the rapid rotation leads to polar
brightening that increases the poles temperature to the hot
side of the jump. At the same time, the rotation leading to gravity
darkening, the equatorial region may be on the cool side of the
jump. Consequently,
the wind in the equatorial region is denser than the
wind in the polar region. Nevertheless, \citet{2000Pelupessy}
show that the rotational velocity of the star should be very
close to its critical value to allow the equatorial wind to reach the
density needed to create the disk. However, supergiant stars
cannot be close to critical rotational velocity because of
probable disruption. Additional mechanisms are therefore
needed to allow the supergiant star to reach its
critical velocity \citep[see
e.g. ][]{2006Owocki}. In the particular case of an sgB[e]
star in an X-ray binary system, the spin-up should occur during the
supergiant phase of the companion, which indicates a different
evolutionary stage from other HMXBs.

This disk itself cannot explain the strong MIR excess the sgB[e] stars
exhibit. Nevertheless, \citet*{1993Bjorkman} have shown
the existence of a zone in the disk (about 50-60 stellar radii
from the star) in which the temperature is below the
temperature of sublimation of the dust (about 1500~K) and the density
high enough to allow for its creation.
\newline

IGR~J16318-4848 is the source in our sample that 
exhibits the strongest MIR excess, and we believe it is due to
the sgB[e] nature of its companion star. Indeed, many other
strongly absorbed sources in our sample do not present any MIR
excess. 

Moreover, it is suggested that IGR~J16358-4726, the second
source in our sample that exhibits a MIR
excess and whose SED needs an additional component to be properly
fitted is an sgB[e], because its spectrum has all the
characteristic features of supergiant stars plus the [\ion{Fe}{ii}]
feature (CHA08). Our fit is therefore in good
agreement with their result, and the only other source of our
sample that definitely exhibits a MIR excess is indeed an sgB[e]
star.

In the case of IGR~J16195-4945, we are more
cautious concerning the presence of warm dust that could be
responsible for a MIR excess, as we lack data above 8
$\mu$m. Indeed, Fig 4. shows that these source could exhibit
a MIR excess, but one much lower than the other two. Nevertheless,
if this excess were to be confirmed, we
believe it would be also due to the sgB[e] nature of the
companion.
\newline

We would like to point out that, because the dust is
most located in an equatorial disk in an
sgB[e] star, the simple model we used to fit the
SEDs cannot reproduce the complex distribution of
the dust around these stars. Nevertheless, it allows the
detection of a warm MIR
excess because of the presence of dust in the stellar
winds. Finally, for all the stars in our sample, we cannot
exclude the presence of a cold
component - responsible for their intrinsic absorption - which we
cannot detect because of the lack of data above 20 $\mu$m.

\subsection{Spectral type and distance}

\begin{table}
	\caption{Summary of spectral types (SpT) and distances
	($\textrm{D}_\ast$) derived from our fits for confirmed
	supergiant stars in our sample. ($^\ast$)
	sources with an accurate spectral type found in the literature
	($^\dagger$) confirmed supergiant stars whose
temperature derived from
	our fits was used to assess their accurate spectral type. 
References to the determination of the spectral type and/or
spectral class of these sources are found in Table 1.}
	$$
	\begin{array}{c c c}
	\hline
	\hline
	\textrm{Sources}&\textrm{SpT}&\textrm{D}_\ast(\textrm{kpc})\\
	\hline
	\textrm{IGR~J16318-4848}^\ast&\textrm{sgB[e]}&\sim1.6\\
	\hline
	\textrm{IGR~J16320-4751}^\dagger&\textrm{O8I}&\sim3.5\\
	\hline
	\textrm{IGR~J16358-4726}^\ast&\textrm{sgB[e]}&\sim18.5\\
	\hline
	\textrm{IGR~J16465-4507}^\ast&\textrm{B0.5I}&\sim9.4\\
	\hline
	\textrm{IGR~J16479-4514}^\dagger&\textrm{O8.5I}&\sim4.9\\
	\hline
	\textrm{IGR~J17252-3616}^\dagger&\textrm{O8.5I}&\sim6.1\\
	\hline
	\textrm{IGR~J17391-3021}^\ast&\textrm{O8I}&\sim2.7\\
	\hline
	\textrm{IGR~J17544-2619}^\ast&\textrm{O9I}&\sim3.6\\
	\hline
	\textrm{IGR~J19140+0951}^\ast&\textrm{B1I}&\sim3.1\\
	\hline
	\end{array}
	$$
\end{table} 

\begin{table}
	\caption{Summary of the distances
	($\textrm{D}_\ast$) derived from our fits for sources
	with unconfirmed spectral classes.}
	$$ 
	\begin{array}{c c ccc}
	\hline
	\hline
	\textrm{Sources}&\textrm{SpT}&\multicolumn{3}{c}{\textrm{D}_
\ast(\textrm{kpc})}\\
	\hline
	&&\textrm{V}&\textrm{III}&\textrm{I}\\
	\hline
	\textrm{IGR~J16195-4945}&\textrm{B1}&\sim3.1&\sim5.7&\sim9.8\\
	\hline
	\textrm{IGR~J16207-5129}&\textrm{O7.5}&\sim1.8&\sim2.8&\sim4.1\\
	\hline
	\textrm{IGR~J16418-4532}&\textrm{O8.5}&\sim4.9&\sim8.3&\sim13\\
	\hline
	\end{array}
	$$ 
\end{table}

In our sample, six sources are supergiant stars with a known
spectral type - IGR~J16318-4848 and IGR~J16358-4726 are sgB[e],
IGR~J16465-4507 is a B0.5I, IGR~J17391-3021 is an 08Iab(f),
IGR~J17544-2619 is an O9Ib, and IGR J19140+0951 is a B1I - and
three are found to be O/B supergiants whose temperatures derived
from our fits allow an assessment of the spectral types
using the classification given in \citet{2005Martins} and
\citet{2006Crowther} for O and B galactic
supergiants, respectively, given the uncertainties of
observational results
($\sim$ 2000~K) and uncertainties on the fits temperatures as given in
Table 6. We therefore found that IGR~J16320-4751 could be an
08I and IGR~J16479-4514 and
IGR~J17252-3616 could be 08.5I stars. 
\newline

Concerning the last three sources whose spectral class is unknown,
results of the fits listed in
Tables 5 and 6 show that they are probably all O/B massive
stars, and we also used their derived temperatures to 
assess their spectral type using the
classification given in both papers quoted
above. IGR~J16195-4945 could be a B1 star, and as already stressed
above, it could be an sgB[e] due to its MIR excess, IGR~J16207-5129
and IGR~J16418-4532 could be O7.5-O8.5 stars. 
Nevertheless, even if the high intrinsic X-ray absorption of
their associated compact objects points towards a supergiant
nature since the
accretion is likely to be wind-fed, the fits
themselves do not allow an assessment of their spectral
classes. We then consider they could be either main
sequence, giant, or supergiant stars.
\newline

\citet*{2006Martins} give a UBVJHK synthetic
photometry of galactic OI, OIII, and OV stars, with which one can
get the expected unabsorbed absolute magnitude in J band
$\textrm{M}_{\rm J}$ for stars having a given spectral
classification. Using the absorbed apparent magnitudes
$\textrm{m}_{\rm J}$
of our sources and the J band absorption we derived
from our fits,
$\textrm{A}_{\rm J}\,=\,0.289\times\textrm{A}_\textrm{v}$, it
is then possible to assess the distance of O stars
in our sample using the standard relation:

\begin{displaymath}
	\,\,\,\,\,\,\,\,\,\,\,\,\,\,\,\,\,\,\,\,\,\,\,\,\,\,\,\,\,\,\,\,\,\,D_
\ast\,=\,10^{\textrm{\normalsize
$0.2(m_{\rm J}-A_{\rm J}-M_{\rm J}+5)$}}\,\,\,\,\,\,\textrm{in
pc}
\end{displaymath}
We did not find any synthetic
photometry for B supergiants. Nevertheless, expected radii of
galactic BI, BIII, and BV stars are given in \citet{1996Vacca},
and we divided these
values by the $\frac{\textrm{R}_\ast}{\textrm{D}_{\ast}}$ ratio
derived from our fits to get the star distance. The derived
distances for the sources whose spectral
class is known are listed in Table 7, in Table 8 for the others.

\subsection{X-ray properties}

Except in the case of IGR~J17544-2619, X-ray absorptions
are systematically significantly larger than the visible
absorptions. This 
indicates the presence in the system of a two-component
absorbing material: one 
around the companion star, responsible for the visible
absorption, and a very dense one around the compact object
coming from the stellar winds that
accrete onto the compact object and are responsible for the huge
X-ray absorption those sources exhibit. 

The obscuration of the compact object by the stellar wind is
caused by the photoelectric absorption of the X-ray emission by the
wind, and this absorption varies along the
orbit of the compact object. This orbital dependence has for
instance been observed and modelled on 4U
1700-37 \citep*{1989Haberl}. 

Moreover, an effect on the X-ray absorption by the photoionisation
of the stellar wind in the vicinity of the compact object by its
X-ray emission was predicted by \citet*{1977Hatchett}. 
Indeed, in SGXBs, the compact object moves through the
stellar wind of the companion star, and the X-rays are
responsible for the enhancement or the depletion of the ionised
atoms responsible
for the acceleration of the wind
(e.g. \ion{C}{iv} and \ion{N}{v}). 
This has a direct consequence on the velocity profile of the
wind; when the wind enters into an ionised zone, it
follows a standard CAK law until it reaches a location in which it
is enough ionised for no further radiative driving to take
place, and the wind velocity is ``frozen'' to a constant value
from this point. This results in a lower wind
velocity close to the compact object and consequently a higher
wind density that leads to a higher obscuration of the compact object.
\newline

Most of the sources studied in this
work are very absorbed in the high-energy
domain. Nevertheless, this absorption may not be always that
high. In
the case of very wide eccentric orbits, the column density of
the sources could
normally vary along their orbit and
suddenly increase when very close to the companion star because
of the wind ionisation. In contrast, if these objects were to
be always very absorbed, it could mean that their orbit is 
very close to the companion star and weakly eccentric. If this
effect were to be observed, we think
it could explain the difference in behaviour between
obscured SGXBs (close quasi-circular orbits) and SFXTs (wide
eccentric orbits).

\subsection{Optical properties}
\begin{table}
	\caption{Sample of parameters we used to fit the SEDs of the
	isolated supergiants. We give their
	galactic coordinates, their spectral types, the interstellar
	extinction in magnitudes $\textrm{A}_\textrm{i}$ and then the
	parameters themselves: the extinction in the optical
	$\textrm{A}_\textrm{v}$, the temperature
	$\textrm{T}_\ast$ and the
	$\frac{\textrm{R}_{\ast}}{\textrm{D}_{\ast}}$ ratio of the
	star.}
	$$
	\begin{array}{c c c c c c c c c c}
	\hline
	\hline
	\textrm{Sources}&l&b&\textrm{SpT}&\textrm{A}_\textrm{i}&\textrm{A}
_\textrm{v}&\textrm{T}_{\ast}
(\textrm{K})&\frac{\textrm{R}_{\ast}}{\textrm{D}_{\ast}}\\
	\hline
	\textrm{HD~144969}&333.18&2.0&\textrm{B0.5Ia}&3.34&3.9&26000&4.01
\times10^{-{10}}\\
	\hline
	\textrm{HD~148422}&329.92&-5.6&\textrm{B0.5Ib}&0.75&0.9&24700&8.76
\times10^{-{11}}\\
	\hline
	\textrm{HD~149038}&339.38&2.51&\textrm{B1Ia}&0.81&1&24000&5.30
\times10^{-{10}}\\
	\hline
	\textrm{HD~151804}&343.62&1.94&\textrm{O8Iaf}&0.83&1.3&32000&4.20
\times10^{-{10}}\\
	\hline
	\textrm{HD~152234}&343.46&1.22&\textrm{B0.5Ia}&1.17&1.5&25100&4.94
\times10^{-{10}}\\
	\hline
	\textrm{HD~152235}&343.31&1.1&\textrm{B1Ia}&3.37&3.9&24500&1.13
\times10^{-{9}}\\
	\hline
	\textrm{HD~152249}&343.35&1.16&\textrm{O9Ib}&1.34&1.7&30100&2.89
\times10^{-{10}}\\
	\hline
	\textrm{HD~156201}&351.51&1.49&\textrm{B0.5Ia}&2.68&2.9&26500&2.90
\times10^{-{10}}\\
	\hline
	\end{array}
	$$
\end{table} 

We were able to fit all but three sources
with a simple stellar black body model. For these three
sources, we explained that the MIR excess was probably caused by
the warm dust created within the stellar wind due to the
sgB[e] nature of the companions. Therefore, it seems that the
optical-to-MIR wavelength emission of these SGXBs corresponds
to the emission of absorbed blue supergiants or sgB[e]. 

Moreover, the results of the fits listed in the Table 5 show
that it is \textit{a priori} impossible to
differentiate an obscured SGXB and an SFXT from their
optical-to-MIR wavelength SEDs, and it then seems that the difference in
behaviour between both kinds of SGXBs only depends on the
geometry of the system,
i.e. its orbital distance or its orbit eccentricity \citep*{2006Chaty}. 
\newline

Nevertheless, to assess a possible effect of the
compact object on the companion star, we took a sample of eight
isolated O/B supergiants in the direction of the Galactic
centre and fitted their optical-to-NIR wavelength SEDs with an absorbed
stellar black body. The best-fitting parameters are listed in Table
9 along to their galactic coordinates, their spectral types and
the interstellar \ion{H}{i} absorption (A$_\textrm{i}$). The
distances of these supergiants are known, which allowed
us to calculate A$_\textrm{i}$ out to their
position using the tool available on the MAST website
\citep{1994Fruscione}.

We see that their visible absorption is
of the same order of 
magnitude as the interstellar \ion{H}{i} absorption and well
below the level of absorption of our sources. This could mean that 
some supergiant stars in SGXBs exhibit an excess of absorption due to
a local absorbing component. Unfortunately, the total
interstellar absorption out to the distance
of our sources is unknown, and we cannot compare
their visible absorptions derived from our fits to the total
interstellar absorption out to their position.

Nevertheless, if this was the case, we think that this excess of 
absorption could also be caused partly by the photoionisation of the
wind in the vicinity of the companion star by the high-energy
emission of the compact object, as this would make their winds
slower than in isolated supergiant stars. Since the wind velocity
is lower, the medium is denser and suitable for creating a more
absorbant material.

Indeed, in the case of persistent sources with very
close and quasi-circular orbits, we think that this
possible effect could be particularly strong, since the wind around
the companion star
would be permanently photoionised and would have lower
velocities than in isolated supergiants. This could be the
general scheme of obscured SGXBs.

On the other hand, in the case of very wide and eccentric
orbits, the compact object would be most of the time far from the
secondary and its X-ray emission would not photoionise
the wind close to the companion star, which would not exhibit any
visible absorption excess until the compact object got
closer. This could be the general scheme of SFXTs.

At last, in both cases, it would be possible to observe a
variation in the P-Cygni profiles of the companion star
(i.e. a variation in the wind velocity) with the phase angle of
the compact object along its orbit.
\newline

As a possible confirmation of this general behaviour, we point
out that the visible absorptions derived from our fits for 
the companion stars of the only sources in our sample that
surely exhibit the SFXT
behaviour (IGR~J16465-4945, IGR~J17391-3021, and IGR~J17544-2619)
are far smaller than the visible absorptions of the
others. Moreover, concerning obscured SGXBs, the wind velocity
of IGR~J16318-4848 was found to be $\sim$ 410~km~s$^{-{1}}$
\citep*{2004Filliatre}, far lower than
the expected wind velocity for O/B supergiants ($\sim$ 1000-2000~km~s$^{-{1}}$).

\section{Conclusions}

In this paper, we presented results of observations performed
at ESO/VLT with VISIR, which aimed at studying the MIR emission of twelve
\textit{INTEGRAL} obscured HMXBs, whose
companions are confirmed or candidate supergiants. Moreover,
using the observations performed at ESO/NTT and reported in the
companion paper (CHA08), previous optical/NIR observations found
in the literature and archival data
from USNO, 2MASS, and GLIMPSE, we fitted the broad-band SEDs of
these sources using a simple two-component black body model in
order to obtain their visible absorptions and temperatures, and
to assess the contribution of their enshrouding material in their
emission. 

We confirmed that all these sources were likely O/B supergiant
stars and that,
for most of them, the enshrouding material marginally contributed to the
emission. Moreover, in the case of IGR~J16318-4848,
IGR~J16358-4726, and perhaps
IGR~J16195-4945, the MIR excess
could be explained by the sgB[e] nature of the companion stars. 

By comparing the optical and high-energy characteristics of
these sources, we showed that the distinction SFXTs/obscured
SGXBs does not seem to exist from optical-to-MIR
wavelength. 
Nevertheless, most of the sources in our sample are
significantly absorbed in the optical, and we think that the wind
can be denser
around some supergiants in SGXBs, which could be due to the
photoionisation by the high-energy emission 
of the compact object.
\newline

Several improvements in our study are needed to
allow definitive conclusions. Indeed, the data used to perform
the SEDs were not taken simultaneously, which can for instance
lead to an incorrect assessment of the MIR excess in the emission.
Moreover, the lack of optical magnitudes for several sources
could have led to an incorrect fitting of their intrinsic visible
absorption $\textrm{A}_\textrm{v}$.
Finally, the absence of an accurate measurement of the total
interstellar absorption out to the distance of these sources
does not allow us to say whether the
presence of the compact object can lead to a stellar wind denser
in some supergiants belonging to SGXBs than in isolated supergiants.
\newline

We then recommend further
optical investigations of these sources to study any
possible variation in their P-cygni profile 
with the phase of the compact object. We also think that the
measurement of the distance of these
sources is crucial to allow a good assessment of the real
interstellar absorption up to their distance, in order to detect
any local absorbing component around companion stars. Finally,
we recommend X-ray monitoring so as to study the
dependence of their column density on orbital phase angle,
which could help for understanding the difference between obscured
SGXBs and SFXTs.

\begin{acknowledgements}
	
	We are pleased to thank J\'er\^ome Rodriguez for his very
	useful website in which all the \textit{INTEGRAL} sources are
	referenced (
	http://isdc.unige.ch/$\sim$rodrigue/html/igrsources.html).
	\newline
	Based on observations carried out at the European Southern
	Observatory, Chile (through programmes ID. 075.D-0773 and
	077.D-0721). This research has made use of NASA's Astrophysics
	Data System, of the SIMBAD and VizieR databases operated at
	the CDS, Strasbourg, France, of products from the US Naval
	Observatory catalogues, of products from the Two
	Micron All Sky Survey, which is a
	joint project of the University of Massachusetts and the
	Infrared Processing and Analysis Center/California Institute
	of Technology, funded by the National Aeronautics and Space
	Administration and the National Science Foundation as well as
	products from the Galactic Legacy Infrared Mid-Plane Survey
	Extraordinaire, which is a \textit{Spitzer Space Telescope}
	Legacy Science Program.
\end{acknowledgements}
\bibliographystyle{aa}
\bibliography{./mybib}{}
$\,\,\,\,\,\,\,\,\,\,\,\,\,\,\,\,\,\,\,\,\,\,\,\,\,\,\,\,\,\,\,\,\,\,\,\,\,\,\,\,\,\,
\,\,\,\,\,\,\,\,\,\,\,\,\,\,\,\,\,\,\,\,\,\,\,\,\,\,\,\,\,\,\,\,\,\,\,\,\,\,\,\,\,\,
\,\,\,\,\,\,\,\,\,\,\,\,\,\,\,\,\,\,\,\,\,\,\,\,\,\,\,\,\,\,\,\,\,\,\,\,$

Note added in proof: After the submission of this paper,
\citet{2007Moon} reported MIR spectroscopy of IGR J16318-4848, confirming
the presence of two components of warm and cold dust around the
supergiant B[e] companion star. Moreover, \citet{2007Hannikainen}
reported a more accurate B0.5Ia spectral type of IGR~J19140+0951.

\end{document}